\documentclass[aps,pra,twocolumn,graphicx,floatfix]{revtex4}
\usepackage{graphicx,amsmath,amsfonts}
\newcommand{\greeksym}[1]{{\usefont{U}{psy}{m}{n}#1}}
\newcommand{\umu}{\mbox{\greeksym{m}}}

\newcommand{\0}{|0\rangle}
\newcommand{\1}{|1\rangle}

\newcommand{\Yb}{$^{171}$Yb$^+$}

\begin{document}
\title{Simultaneous cooling of axial vibrational modes in a linear ion-trap}
\author{Christof Wunderlich\footnote{Present address: Fachbereich Physik, Universit\"at Siegen,
Walter-Flex-Str.3, 57068 Siegen, Germany}} \affiliation{National
University of Ireland, Maynooth, Maynooth Co.~Kildare, Ireland}
\author{Giovanna Morigi}
\affiliation{Abteilung Quantenphysik, Albert-Einstein-Allee 11,
D-89069 Ulm, Germany} \affiliation{Departament de Fisica,
Universitat Autonoma de Barcelona, E-08193 Barcelona, Spain}
\author{Dirk Rei{\ss}}
\affiliation{Institut f\"ur Laser-Physik, Universit\"at Hamburg,
Luruper Chaussee 146, 22761 Hamburg, Germany}

\date{\today}

\begin{abstract}
In order to use  a collection of trapped ions for experiments where
a well defined preparation of vibrational states is necessary, all
vibrational modes have to be cooled to ensure precise and repeatable
manipulation of the ions' quantum states. A method for simultaneous
sideband cooling of all axial vibrational modes is proposed. By
application of a magnetic field gradient the absorption spectrum of
each ion is modified such that sideband resonances of different
vibrational modes coincide. The ion string is then irradiated with
monochromatic electromagnetic radiation, in the optical or microwave
regime, for sideband excitation. This cooling scheme is investigated
in detailed numerical studies. Its application for initializing ion
strings for quantum information processing is extensively discussed.
\end{abstract}


\maketitle

\graphicspath{{Bilder/}}

\section{Introduction}

Atomic ions trapped in an electrodynamic cage allow for preparation
and measurement of individual quantum systems, and represent an
ideal system to investigate fundamental questions of quantum
physics, for instance, related to decoherence \cite{Myatt00,Roos99},
the measurement process \cite{Hannemann02,Wunderlich03}, or
multiparticle entanglement \cite{Entangle}. Also, trapped ions
satisfy all criteria necessary for quantum computing. Two internal
states of each ion represent one elementary quantum mechanical unit
of information (a qubit). The quantized vibrational motion of the
ions (the ``bus-qubit'') is used as means of communication between
individual qubits to implement conditional quantum dynamics with two
or more qubits \cite{Cirac95}. In recent experiments quantum logic
operations with two trapped ions were realized~\cite{QGate} and
teleportation of an atomic state has been
demonstrated~\cite{Teleport}.

These implementations of quantum information processing (QIP) with
trapped ions require that the ion string is cooled to low
vibrational collective excitations
\cite{Cirac95,QGate,Sorensen00,Jonathan00}. In particular, this
condition should be fulfilled by {\it all} collective vibrational
modes~\cite{Wineland98}. Therefore, in view of the issue of scalable
QIP with ion traps, it is important to find efficient cooling
schemes that allow to prepare vibrationally cold ion chains.

Cooling of the vibrational motion of two ions in a common trap
potential has been demonstrated
experimentally~\cite{King98,Peik99,Roos00,Reiss02} (see also
\cite{Eschner03} for a recent review). This is deemed to be
sufficient for a quantum information processor which utilizes two
ions {\em at a time} for quantum logic operations with additional
ions stored in spatially separated regions \cite{Kielpinski02}. If
more than two ions reside in a common trap potential and shall be
used simultaneously for quantum logic operations, however, the
task of reducing the ions' motional thermal excitation becomes
increasingly challenging with a growing number of ions and
represents a severe obstacle on the way towards scalable QIP with
an ion chain. Straightforward extensions of laser-cooling schemes
for one particle to many ions, like sequentially
applying sideband cooling~\cite{Eschner03} to each one of the
modes, becomes inefficient as the number of ions increases, since
after having cooled the last mode, the first one may already be
considerably affected by heating due to photon scattering and/or
due to fluctuations of the trap potential. Therefore, it is
desirable to find a method that allows for simultaneous and
efficient cooling of many vibrational modes of a chain of ions.

In this article we propose a scheme that allows for simultaneous
sideband cooling of all collective modes of an ion chain to the
ground state. This is achieved by inducing position dependent Zeeman
shifts through a suitably designed magnetic field, thereby shifting
the spectrum of each ion in such a way that the red-sideband
transitions of each mode may occur at the same frequency. Thus, by
irradiating the ion string with monochromatic radiation all axial
modes are cooled. We investigate numerically the efficiency and
explore implementations of simultaneous sideband excitation by means
of laser light, and alternatively, by using long-wavelength
radiation in the radio-frequency or microwave
regime~\cite{Mintert01,Wunderlich02,McHugh05}.

The remainder of this article is organized as follows: In section
\ref{IntroScheme} the cooling scheme is outlined. Numerical
investigations of the cooling efficiency are presented for
implementations using an optical Raman transition (section
\ref{Raman}) and a microwave transition (section \ref{MW}). In
section \ref{Exp} possible experimental implementations are
discussed and the cooling scheme is studied under imperfect
experimental conditions. The paper is concluded in section
\ref{Conclusion}.

\section{The concept of simultaneous sideband cooling}
\label{IntroScheme}

\subsection{Axial vibrational modes}

We consider $N$ crystallized ions each of mass $m$ and charge $e$ in a harmonic
trap. The trap potential has cylindrical symmetry around the $z-$axis providing
strong radial confinement such that the ions are aligned along this
axis~\cite{Footnote:Example}. We denote by $\nu_r$, $\nu_z$ the radial and
axial frequencies of the resulting harmonic potential, where $\nu_r \gg \nu_z$,
and by $z_j^{(0)}$ the ions classical equilibrium positions along the trap
axis. The typical axial distance $\delta z$ between
neighboring ions scales like $\delta z\sim \zeta_0 2 N^{-0.57}$ with
$\zeta_0\equiv(e^2/(4\pi\epsilon_0 m \nu_1^2))^{1/3}$\cite{Steane97,James98}.
For brevity, in the remainder of this article, the ion at the classical
equilibrium position $z_j^{(0)}$ is often referred to as "ion $j$".

At sufficiently low temperatures the ions vibrations around their
respective equilibrium positions are harmonic and the axial
motion is described by $N$ harmonic oscillators according to the
Hamiltonian
\begin{equation}
\tilde{H}_{\rm mec} =  \sum_{\alpha=1}^N \hbar\nu_{\alpha} (a_{\alpha}^\dagger a_{\alpha}+1/2) \
, \label{HamOsz}
\end{equation}
where $\nu_{\alpha}$ are the frequencies of the chain collective modes and
$a_{\alpha}^\dagger$ and $a_{\alpha}$ the creation and annihilation operators
of a phonon at energy $\hbar\nu_{\alpha}$. We denote with
$Q_{\alpha},P_{\alpha}$ the corresponding quadratures, such that
$[Q_{\alpha},P_{\alpha}]={\rm i}\hbar$, and choose the labelling convention
$\nu_1<\nu_2<\ldots<\nu_N$, whereby $\nu_1=\nu_z$ (in this article we often
refer to the collective vibrational mode characterized by $\nu_\alpha$ as "mode
$\alpha$"). The local displacement $q_j=z_j-z_j^{(0)}$ of the ion $j$ from
equilibrium is related to the coordinates $Q_{\alpha}$ by the transformation
\begin{equation}
q_j=\sum_{\alpha}S_j^{\alpha}Q_{\alpha}
\end{equation}
where $S_{j}^{\alpha}$ are the elements of the unitary matrix $S$
that transforms the dynamical matrix $A$, characterizing the ions
potential, such that $S^{-1}AS$ is diagonal. The frequencies,
$\nu_\alpha$ of the vibrational modes are given by
$\sqrt{\upsilon_\alpha} \times \nu_1$ where $\upsilon_\alpha$ are
the eigenvalues of $A$~\cite{James98}. The normal modes are excited
by displacing an ion from its equilibrium position $z_j^{(0)}$ by an
amount $q_j$. Thus, the coefficients $S_{j}^{\alpha}$ describe the
strength with which a displacement $q_j$ from $z_j^{(0)}$ couples to
the collective mode $\alpha$.

Excitation of a vibrational mode can be achieved through the
mechanical recoil associated with the scattering of photons by the
ions. This excitation is scaled by the Lamb-Dicke parameter (LDP)
\cite{Stenholm86}, which for a single ion corresponds to
$\sqrt{\omega_R/\nu}$, where $\omega_R=\hbar k^2/2m$ is the recoil
frequency and $\hbar k$ the linear momentum of a photon. In an ion
chain we associate a Lamb-Dicke parameter $\eta_{\alpha}$ with
each mode according to the equation
\begin{equation}
\eta_{\alpha}=\sqrt{\frac{\omega_R}{\nu_{\alpha}}}\ .
\label{eta_alpha}
\end{equation}
Hence, if a photon is scattered by the ion at $z_j^{(0)}$, the
ion recoil couples to the mode $\alpha$ according to the
relation~\cite{Morigi01}
\begin{equation}
\eta_{j}^{\alpha}=S_{j}^{\alpha}\eta_{\alpha} \ .
 \label{eta}
\end{equation}
In the remainder of this article we will assume that the ions are
in the Lamb-Dicke regime, corresponding to the fulfillment of
condition $\sqrt{\langle a_{\alpha}^\dagger a_{\alpha}\rangle}
\eta_{\alpha}
 \ll 1$. In this regime the scattering of a photon does not couple to the
vibrational excitations at leading order in this small parameter, while changes
of one vibrational quantum $\hbar\nu_{\alpha}$  occur with probability that
scales as $|\eta_{\alpha}|^2$. Changes by more than one vibrational quantum are
of higher order and are neglected here.

\subsection{Sideband cooling of an ion chain}
\label{Sec:theorie}

In this section we consider a schematic description of sideband
cooling of an ion chain, in order to introduce the  concepts
relevant for the following discussion. We denote by $\0$ and $\1$
the internal states of the ion transition at frequency $\omega_0$,
in absence of external fields, and linewidth $\gamma$. A spatially
inhomogeneous magnetic field is applied that shifts the transition
frequency of each ion individually such that for the ion at position
$z_j^{(0)}$ the value $\omega_j$ is assumed. Each ion transition
couples to radiation at frequency $\omega_L$, which drives it well
below saturation. In this limit, the contributions of scattering
from each ion to the excitation of the modes add up incoherently
\cite{Morigi99,Morigi01}.

For this system, the equations describing the dynamics of laser
sideband cooling of an ion chain can be reduced to rate equations
of the form
\begin{eqnarray}
\label{Palpha} \frac{\rm d}{{\rm d}t}P_{\alpha}(n^{(\alpha)})
&=&(n^{(\alpha)}+1)\left[A_{-}^{\alpha}P_{\alpha}(n^{(\alpha)}+1)-A_{+}^{\alpha}
P_{\alpha}(n^{(\alpha)})\right]\nonumber\\
& &-n^{(\alpha)}\left[A_{-}^{\alpha}P_{\alpha}(n^{(\alpha)})-
A_{+}^{\alpha}P_{\alpha}(n^{(\alpha)}-1)\right]
\end{eqnarray}
where $P_{\alpha}(n^{(\alpha)})$ is the average occupation of the vibrational number
state $|n^{(\alpha)}\rangle$ of the mode $\alpha$, and $A_{+}^{\alpha}$
($A_{-}^{\alpha}$) characterizes the rate at which the mode is
heated (cooled). Equation~(\ref{Palpha}) is valid in the
Lamb-Dicke regime, i.e.\ when the LDP is sufficiently small to
allow for a perturbative expansion in this parameter. Denoting by
$\Omega_j$ the Rabi frequency, the heating and cooling rate takes
the form~\cite{Morigi01}
\begin{equation}
\label{Aplus}
A_{\pm}^{\alpha}=\sum_{j=1}^N|\eta_{j}^{\alpha}|^2\frac{\Omega_j^2}{2\gamma}
\left[\frac{\gamma^2}{4(\delta_j\mp\nu_{\alpha})^2+\gamma^2}+
\phi\frac{\gamma^2}{4\nu_{\alpha}^2+\gamma^2}\right]
\end{equation}
where the detuning $\delta_j\equiv\omega_L-\omega_j$. The
coefficient $\phi$ emerges from the integral over the angles of
photon emission, according to the pattern of emission of the given
transition \cite{Stenholm86}.
For $A_{-}^{\alpha}>A_{+}^{\alpha}$ a steady state exists, it is approached at
the rate
\begin{equation}
\Gamma_{\rm cool}^{(\alpha)}=A_{-}^{\alpha}-A_{+}^{\alpha}
\label{Gamma}
\end{equation}
and the average number of phonons of mode $\alpha$ at steady state
is given by the expression
\begin{equation}
\label{nsteady}
\langle n^{(\alpha)}\rangle=
  \frac{A_{+}^{\alpha}}{A_{-}^{\alpha}-A_{+}^{\alpha}} \ .
\end{equation}
Sideband cooling reaches $\langle n^{(\alpha)}\rangle\ll 1$ through
$A_{-}^{\alpha}\gg A_{+}^{\alpha}$. This condition is obtained by selectively
addressing the motional resonance at $\omega_0-\nu_{\alpha}$. This is
accomplished for a single collective mode when $\gamma\ll\nu_{\alpha}$ and
$\delta_{\alpha}=\nu_{\alpha}$.

In this work, we show how the application of a suitable magnetic
field allows for simultaneous sideband cooling of all modes. In
particular, the field induces space-dependent frequency shifts
that suitably shape the excitation spectrum of the ions.
Simultaneous cooling is then achieved when for each mode $\alpha$
there is one ion $j$ with the matching resonance frequency, that
is, such that $\delta_j = \omega_L - \omega_j = \nu_{\alpha}$.
This procedure is outlined in detail in the following subsection.

\subsection{Shaping the spectrum of an $N$ ion chain}

\begin{figure}[htbp]
\begin{center}
\includegraphics[width=8cm]{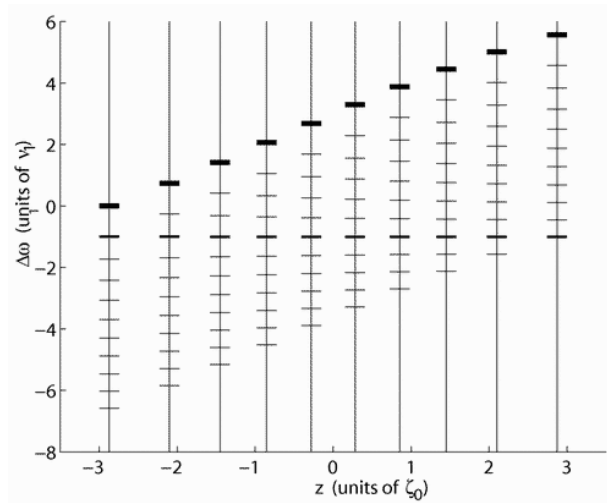}
\caption{Illustration of the axial motional spectrum of a chain of
10 ions in the presence of a spatially inhomogeneous magnetic field.
The vertical lines indicate the axial position of each ion in units
of $\zeta_0$. The corresponding horizontal lines indicate the
frequencies of the spectral lines as measured at that particular
ion. The thick horizontal lines indicate the ions resonance
frequencies $\omega_j - \omega_1$ (in units of the secular axial
frequency $\nu_1$) relative to the resonance frequency $\omega_1$ of
the ion at $z_1=-2.87\zeta_0$. The remaining horizontal lines show
the frequencies of red sideband resonances for each ion at
frequencies $\omega_j -\nu_\alpha$ ($j,\alpha=1,\ldots,10$). The
magnetic field is designed such that $\omega_1-\nu_1 =
\omega_2-\nu_2 = \ldots \omega_N-\nu_N$.  These resonances are
highlighted by medium thick lines.} \label{scheme}
\end{center}
\end{figure}

Assume the ion transition $|0\rangle\to |1\rangle$ and that a
magnetic field--whose magnitude varies as a function of $z$--is
applied to the linear ion trap, Zeeman shifting this resonance. As
a result, the ions resonance frequencies $\omega_j$ are no longer
degenerate. The field gradient is designed such that all ions
share a common motion-induced resonance. This resonance
corresponds to one of the transitions $|0,n^{(\alpha)}\rangle \to
|1,n^{(\alpha)}-1\rangle$, namely to the red sideband of the modes
$\alpha$. The resonance frequency of each ion is shifted such that
the red sidebands of all modes can be resonantly and
simultaneously driven by monochromatic radiation at frequency
$\omega=\omega_1-\nu_1=\ldots =\omega_N-\nu_N$. Ionic resonances
and the associated red sideband resonances--optimally shifted for
simultaneous cooling--are illustrated in Fig.~\ref{scheme} for the
case of 10 ions.

Sideband excitation can be accomplished by either laser light or
microwave radiation according to the scheme discussed
in~\cite{Mintert01}. With appropriate recycling schemes this leads
to sideband cooling on all $N$ modes simultaneously. A discussion on
how a suitable field gradient shifting the ionic resonances in the
desired fashion can be generated is deferred to section~\ref{Exp}.

\subsection{Theoretical model}
\label{Sec:Model}

\begin{figure}[htbp]
\begin{center}
\includegraphics[width=4cm]{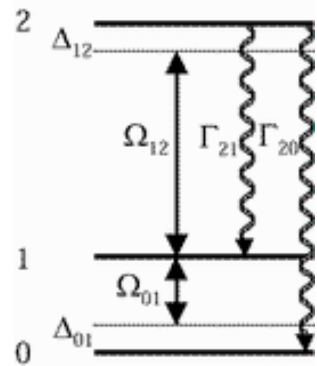}
\caption{Schematic of the ions internal energy levels on which
cooling is implemented. Indicated are the relevant Rabi
frequencies (symbols $\Omega$), spontaneous decay rates
($\Gamma$), and detunings ($\Delta$). The corresponding equations for the
dynamics are discussed in detail in the appendix} \label{levels}
\end{center}
\end{figure}

As an example, we discuss simultaneous sideband cooling of the collective axial
modes of a chain composed of \Yb ions with mass $m=171$ a.m.u.. The ions are
crystallized along the axis of a linear trap characterized by $\nu_1= 1\times
2\pi$MHz. A magnetic field $B(z)$ along the axis is applied that Zeeman-shifts
the energy of the internal states. The value of the field along $z$ is such
that it shifts the red-sidebands of all modes into resonance along the chain,
while at the same time its gradient is sufficiently weak to negligibly affect
the frequencies of the normal modes \cite{Wunderlich02}.

The selective drive of the motional sidebands can be implemented
on a magnetic dipole transition in \Yb close to
$\omega_0=12.6\times 2\pi$ GHz between the hyperfine states
$|0\rangle=|S_{1/2},F=0\rangle$ and
$|1\rangle=|S_{1/2},F=1,m_F=1\rangle$. The magnetic field gradient
lifts the degeneracy between the resonances of individual ions,
and the transition frequency $\omega_j$ of ion $j$ is proportional
to $B(z_j)$ in the weak field limit $\mu_B B/\hbar\omega_0 \ll
1$, where $\mu_B$ is the Bohr magneton. For strong magnetic fields the
variation of $\omega_j$ with $B$ is obtained from the Breit-Rabi
formula \cite{Wunderlich03}.

We investigate two cases, corresponding to two different implementations of the
excitation of the sideband transition between states $\0$ and $\1$. In the
first case, discussed in section \ref{Raman}, the sideband transition is driven by two lasers with appropriate detuning, namely a Raman transition is implemented with intermediate state
$|2\rangle = |P_{1/2}\rangle$. In the second case, presented in section \ref{MW},
microwave radiation drives the magnetic dipole.

Since spontaneous decay from state $\1$ back to $\0$ is negligible on this
hyperfine transition, laser light is used to optically pump the ion into the
$\0$ state via excitation of the $|1\rangle\to|2\rangle$ electric dipole
transition. This laser light is close to 369nm and serves at the same time for state
selective detection by collecting resonance fluorescence on this transition,
and for initial Doppler cooling of the ions. The state $|2\rangle$ decays with
rates $\Gamma_{21}=11 \times 2\pi$MHz and $\Gamma_{20}=5.5\times 2\pi$MHz into
the states $|1\rangle$ and $|0\rangle$, respectively~\cite{FootnoteB}. The
considered level scheme is illustrated in Fig.~\ref{levels}, and the corresponding model
is described in the appendix.

We evaluate the efficiency of the cooling procedure by neglecting the coupling between
different vibrational modes by photon scattering, which is reasonable when the
system is in the Lamb-Dicke regime. In this case, the dynamics reduce to
solving the equations for each mode $\alpha$ independently, and
the contributions from each ion to the dynamics of the mode are summed up incoherently~\cite{Morigi01}, as outlined in Sec.~\ref{Sec:theorie}.
The steady state and cooling rates for each mode are evaluated using the method discussed
in~\cite{Marzoli94} and extended to a chain of $N$ ions. The extension of this method
to a chain of ions is presented in the appendix. The numerical calculations were carried out
for this scheme and chains of $N$ ions with $1<N\leq 10$ and for some values
$N>10$. Since the qualitative conclusions drawn from these calculations did not
depend on $N$, we therefore restrict the discussion in sections
\ref{Raman}, \ref{MW}, and \ref{Exp} to the case $N=10$.

\section{Raman sideband cooling of an ion chain}
\label{Raman}

We consider sideband cooling of an ion chain when the red sideband
transition is driven by a pair of counter-propagating lasers, which
couple resonantly the levels $|0\rangle$ and $|1\rangle$. The two
counter-propagating light fields couple with frequency
$\omega_{R1}$, $\omega_{R2}$ to the optical dipole transitions
$|0\rangle\to |2\rangle$ and $|1\rangle\to |2\rangle$, respectively.
The two lasers are far detuned from the resonance with level
$|2\rangle$ such that spontaneous Raman transitions are negligible
compared to the stimulated process. We denote by
$\Delta_{01}=[(\omega_{R1}-\omega_{R2})-\omega_1]$ the Raman detuning, such that
$\Delta_{01}=0$ corresponds to driving resonantly the transition
$|0\rangle\to |1\rangle$ at the first ion in the chain, and by
$\Omega_{01}$ the Rabi frequency describing the effective coupling
between the two states. A third light field with Rabi frequency
$\Omega_{12}$ is tuned close to the resonance
$|1\rangle\to|2\rangle$ and serves as repumper into state
$|0\rangle$ (compare Fig.~\ref{levels}). The frequencies
$\omega_{Ri}$ are close to the \Yb resonance at 369nm, and the trap
frequency is $\nu=1\times 2\pi$MHz. Hence, from
Eq.~(\ref{eta_alpha}) the Lamb-Dicke parameter takes the value
$\eta_1\approx 0.0926$.

\subsection{Sequential cooling}
\label{SeqCoolRate}

\begin{figure}[htbp]
\begin{center}
\includegraphics[width=6cm]{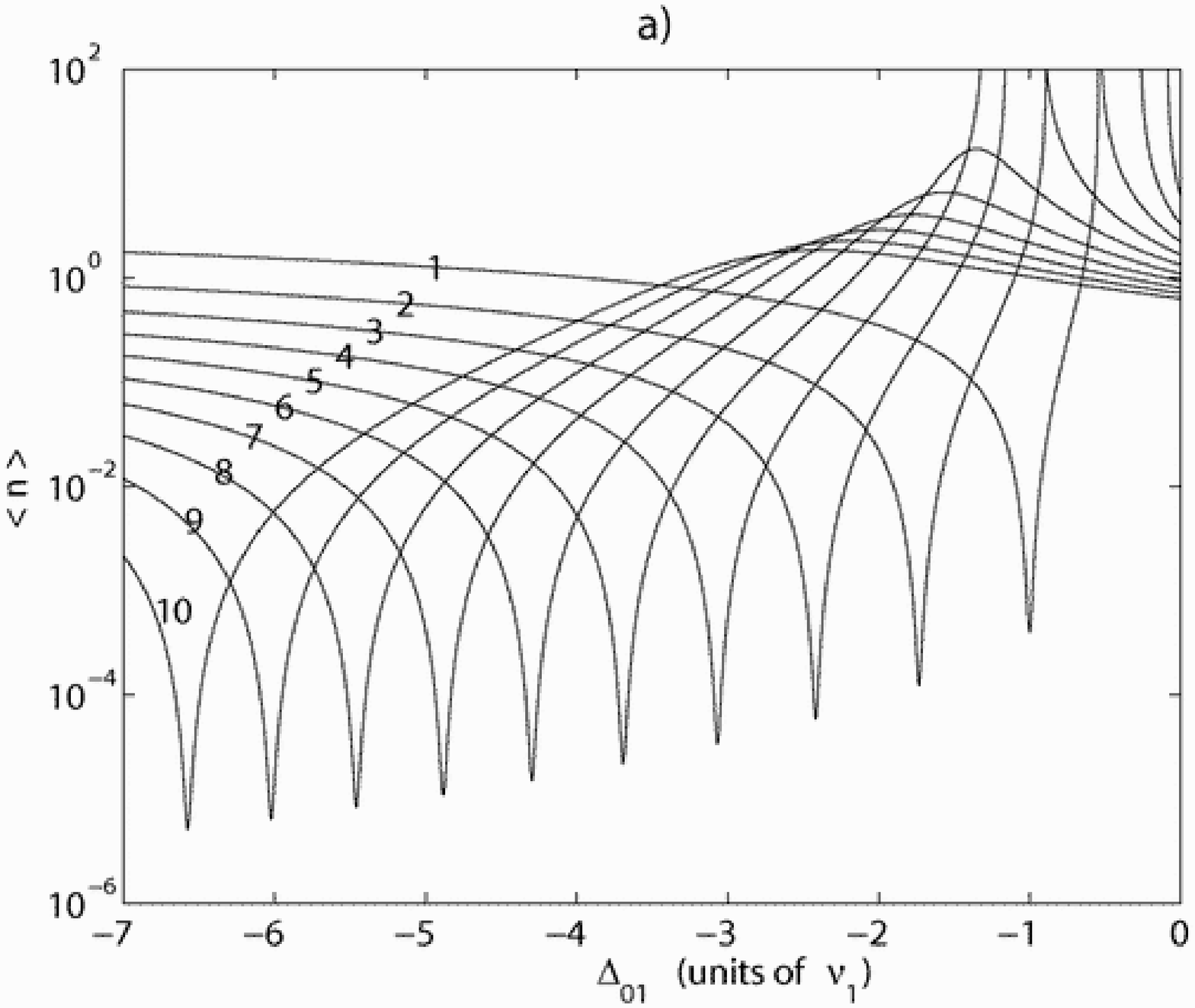}
\includegraphics[width=6cm]{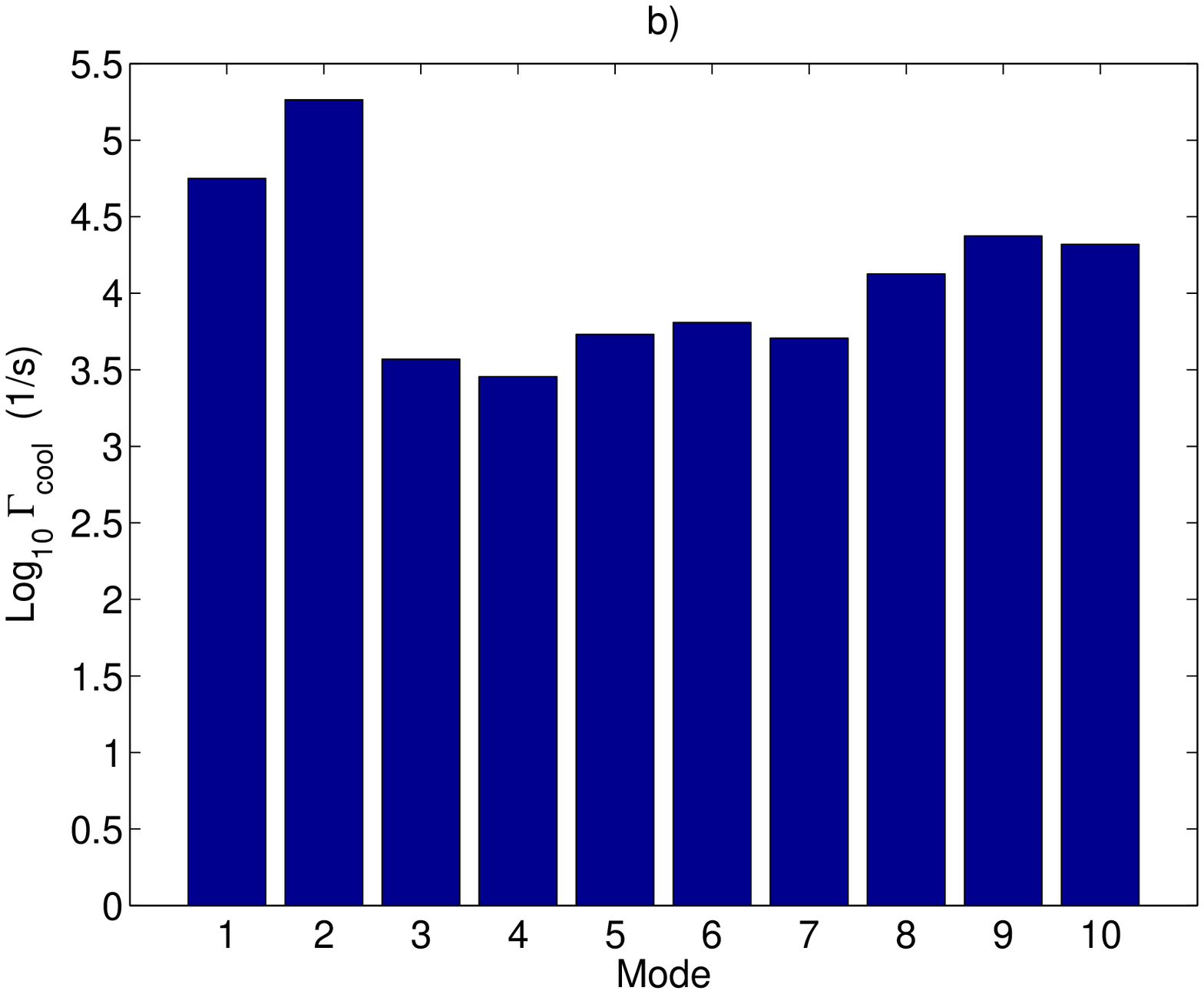}
\includegraphics[width=6cm]{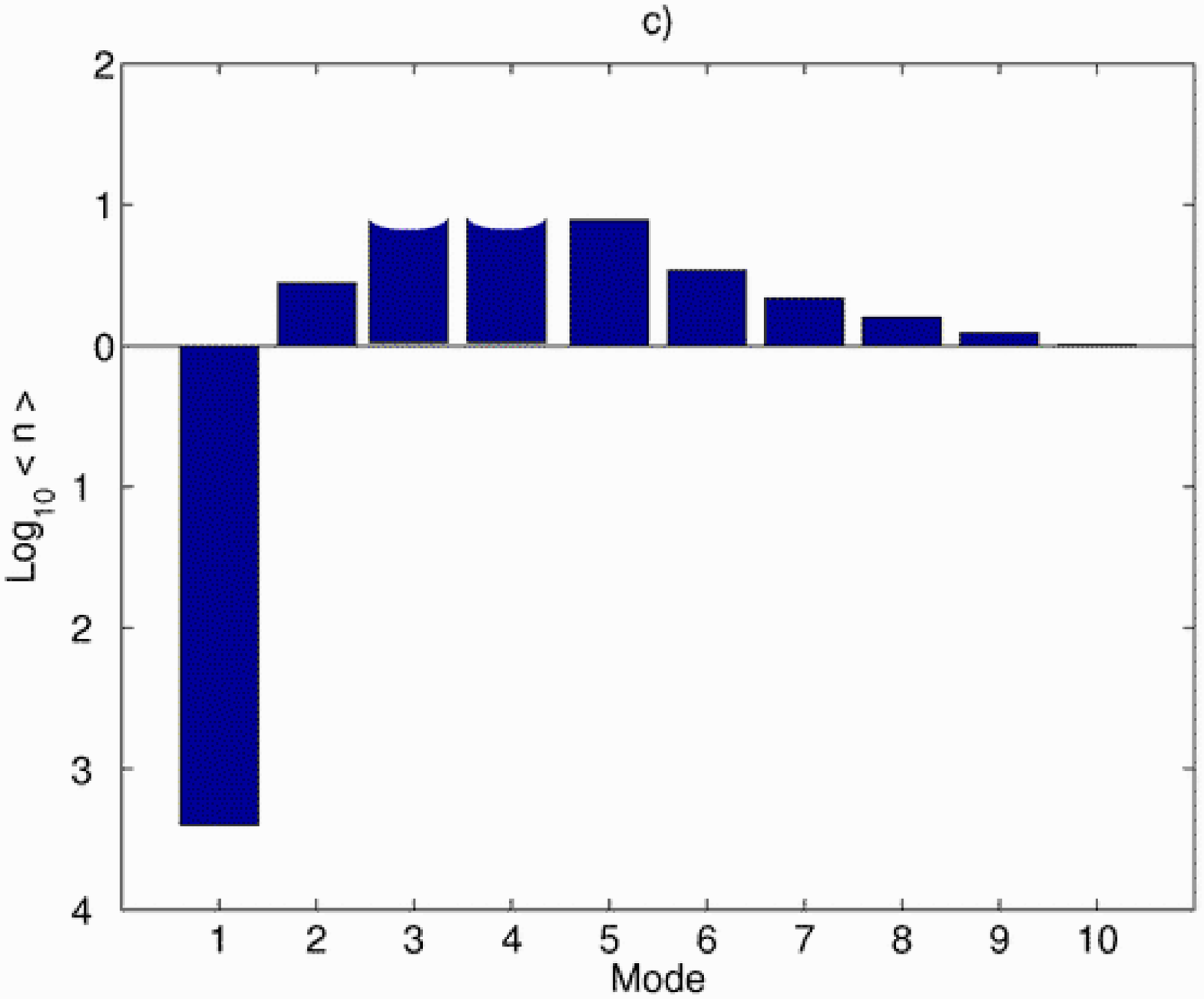}
\caption{Raman sideband cooling of a chain of 10 ions {\em
without} magnetic field gradient. The parameters are
$\Omega_{12}=100\times 2\pi$kHz, $\Omega_{01}=30 \times 2\pi$kHz,
$\Delta_{12} =-10\times 2\pi$MHz. a) Steady state mean vibrational
number $\langle n^{(\alpha)}\rangle$ as a function of
  $\Delta_{01}$ in units of $\nu_1$. b)  Cooling rate $\Gamma_{\rm
cool}^{(\alpha)}$ of mode $\alpha$ for $\Delta_{01}=-\nu_{\alpha}$.
c) Steady state populations for $\Delta_{01}=-\nu_1$, corresponding
to sideband cooling of mode 1. The bars indicating the vibrational
excitations of modes
 $\nu_3$ and $\nu_4$ have been truncated.}
 \label{cool_woB}
\end{center}
\end{figure}

In absence of external field gradients shifting inhomogeneously
the ions transition frequencies (namely, when
$\omega_1=\ldots=\omega_N=\omega_0$), cooling of an ion chain
could be achieved by applying sideband cooling to each mode
sequentially. In each step of the sequence all ions are
illuminated simultaneously by laser light with detuning
$\Delta_{01}=-\nu_{\alpha}$, thereby achieving sideband cooling of
a particular mode $\alpha$. Since all ions are illuminated, they
all contribute to the cooling of mode $\alpha$.

In Fig.~\ref{cool_woB}a) the steady state vibrational number of
each mode at the end of the cooling dynamics is displayed as a
function of the relative detuning $\Delta_{01}$. Each mode
$\nu_{\alpha}$ reaches its minimal excitation at values of the
detuning $\Delta_{01}=-\nu_{\alpha}$. Therefore, in order to cool
all modes close to their ground state, the detuning of the laser
light has to be sequentially set to the optimal value for each
mode $\alpha$.

The cooling rates $\Gamma^{(\alpha)}_{\rm cool}$ of mode $\alpha$ at
$\Delta_{01}= -\nu_\alpha$, as defined in Eq.~(\ref{Gamma}), are
displayed in Fig.~\ref{cool_woB}b). They are different for each mode
and vary between 1kHz and 100kHz for the parameters chosen here.
Even though these cooling rates would, in principle, allow for
cooling sequentially all modes in a reasonably short time, this
scheme may not be effective, since while a particular mode $\alpha$
is cooled all other modes are heated (i) by photon recoil, and, (ii)
by coupling to the environment. As external source of heating we
consider here the coupling of the ions charges to the fluctuating patch
fields at the electrodes~\cite{Turchette00}. The effects of these
processes on the efficiency of cooling are discussed in what
follows.

The consequences of heating due to photon scattering are visible in
Fig.~\ref{cool_woB}a). Here, one can see that while cooling one
mode, others can be simultaneously heated, such that their average
phonon number at steady state is very large. These dynamics are due
to the form of the resonances in a three-level
configuration~\cite{Marzoli94,EIT00}. In general, however, the time
scale of heating processes due to photon scattering is considerably
longer than the time scale at which a certain mode is optimally
sideband cooled, since the transitions leading to heating are out of
resonance. In the case discussed in Fig.~\ref{cool_woB}, for
instance, the heating rates of these modes at $\Delta_{01}=-\nu_1$
are orders of magnitude smaller than the cooling rate of mode 1, and
their dynamics can be thus neglected while mode 1 is sideband
cooled. Similar dynamics are found for $\Delta_{01}=-\nu_{\alpha}$.
Thus, in general one may neglect photon scattering as source of
unwanted heating of modes that are not being efficiently cooled.

Nevertheless, heating by fluctuating electric fields occurs with
appreciable rates ranging between 5s$^{-1}$ and
$10^4$s$^{-1}$~\cite{Turchette00}. The heating rate is different for
each mode and was observed to be considerably larger for the COM
mode (here denoted as mode 1) than for modes that involve
differential relative displacements of individual ions. Obviously,
cooling can only be achieved, if the cooling rate
$\Gamma_{\rm cool,seq}^{(\alpha)}$ of each mode exceeds in magnitude the
corresponding trap heating rate denoted by $\Gamma_{\rm
heat}^{(\alpha)}$:
\begin{equation}
\Gamma_{\rm heat}^{(\alpha)} \ll \Gamma_{\rm cool,seq}^{(\alpha)}
\forall \alpha=1,\ldots ,10 \ .
 \label{CoolHeat}
\end{equation}

In addition, one must consider that {\em after} a particular mode
$\alpha$ has been cooled, it might heat up again while {\em all
other} modes, $\beta\neq \alpha$ are being cooled. This imposes a
second condition on the cooling rate. In order to quantify this
second condition, we first evaluate the time, $T_{\rm
cool}^{(\alpha)}$ it takes to cool {\em one} particular mode
$\alpha$ from an initial thermal distribution, obtained by means of
Doppler cooling and characterized by the average occupation number
$\langle n^{(\alpha)}\rangle_i$, to a final distribution
characterized by $\langle n^{(\alpha)}\rangle_f$. This time can be
estimated to be~\cite{Stenholm86}
\begin{equation}
T_{\rm cool}^{(\alpha)} \equiv \ln \frac{\langle
n^{(\alpha)}\rangle_i} {\langle n^{(\alpha)}\rangle_f} /
(\Gamma_{\rm cool,seq}^{(\alpha)}-\Gamma_{\rm heat}^{(\alpha)}) \ .
\end{equation}
where $\langle n^{(\alpha)}\rangle_f\ll 1$ at steady state was
assumed.

From this relation one obtains the total time, $T_{\rm
seq}^{(\alpha)}$ needed to cool all modes {\em except}
mode $\alpha$, or, in other words the time during which mode
$\alpha$ is not cooled and could get heated. This time, $T_{\rm seq}^{(\alpha)}$
needed to sideband cool all modes with $\beta \neq \alpha$ is
\begin{equation}
T_{\rm seq}^{(\alpha)}=\sum_{\beta,\beta \neq \alpha} T_{\rm cool}^{(\beta)}
 =  \sum_{\beta \neq \alpha}\ln \frac{\langle n^{(\beta)}\rangle_i} {\langle
n^{(\beta)}\rangle_f} 1 /
 (\Gamma_{\rm cool,seq}^{(\beta)}-\Gamma_{\rm heat}^{(\beta)}) \ .
\label{T_heat_seq}
\end{equation}
If mode $\alpha$ is to stay cold during this time, the heating rate
affecting it must be small enough. Hence the condition for efficient
sequential sideband cooling is derived,
\begin{equation}
\Gamma_{\rm heat}^{(\alpha)} \times T_{\rm seq}^{(\alpha)} \ll 1
\label{SeqCool}
\end{equation}
namely, during time $T_{\rm seq}^{(\alpha)}$, necessary for cooling
the modes $\beta\neq \alpha$, the heating of mode $\alpha$ has to be
negligible. Clearly, this condition is stronger than the one derived
in relation (\ref{CoolHeat}), and its fulfillment becomes critical
as the number of vibrational modes (ions) is increased.

A rough estimate of the time $T_{\rm seq}^{(\alpha)}$ to be inserted
in~(\ref{SeqCool}) can be obtained from eq.~(\ref{T_heat_seq}) under
the assumption that all modes start out with the same mean
excitation $\langle n\rangle_i$ (usually determined by initial
Doppler cooling) and are cooled to the same final excitation
$\langle n\rangle_f$. Using condition~(\ref{CoolHeat}), one obtains
\begin{equation}
T_{\rm seq}^{(\alpha)} \approx \ln \frac{\langle n\rangle_i}
{\langle n \rangle_f} \sum_{\beta \neq \alpha} 1 /
 \Gamma_{\rm cool,seq}^{(\beta)} \ .
\label{T_heat_seq_app}
\end{equation}
Substituting this expression into~(\ref{SeqCool}) gives
\begin{equation}
\Gamma_{\rm heat}^{(\alpha)} \ll \left(\ln \frac{\langle n\rangle_i}
{\langle n \rangle_f} \sum_{\beta \neq \alpha} 1 /\Gamma_{\rm
cool,seq}^{(\beta)}\right)^{-1} \equiv \Gamma_{\rm <,seq}
 \label{G_heat}
\end{equation}
which places a stronger restriction than~(\ref{CoolHeat}) on the
trap heating rate that can be tolerated, if sequential cooling is to
work. This relation has to hold true for $\alpha=1,\ldots ,N$.
Expression (\ref{G_heat}) will be used for a  comparison with
simultaneous sideband cooling (see Sec.~\ref{Sec:Compare}).

\subsection{Simultaneous cooling}
\label{Optical}

We consider now the case, when a magnetic field gradient is applied
to the ion chain, such that the situation shown in Fig.~\ref{scheme}
is realized. The axial modes of the chain can then be
simultaneously cooled.

\begin{figure}[htbp]
\begin{center}
\includegraphics[width=6cm]{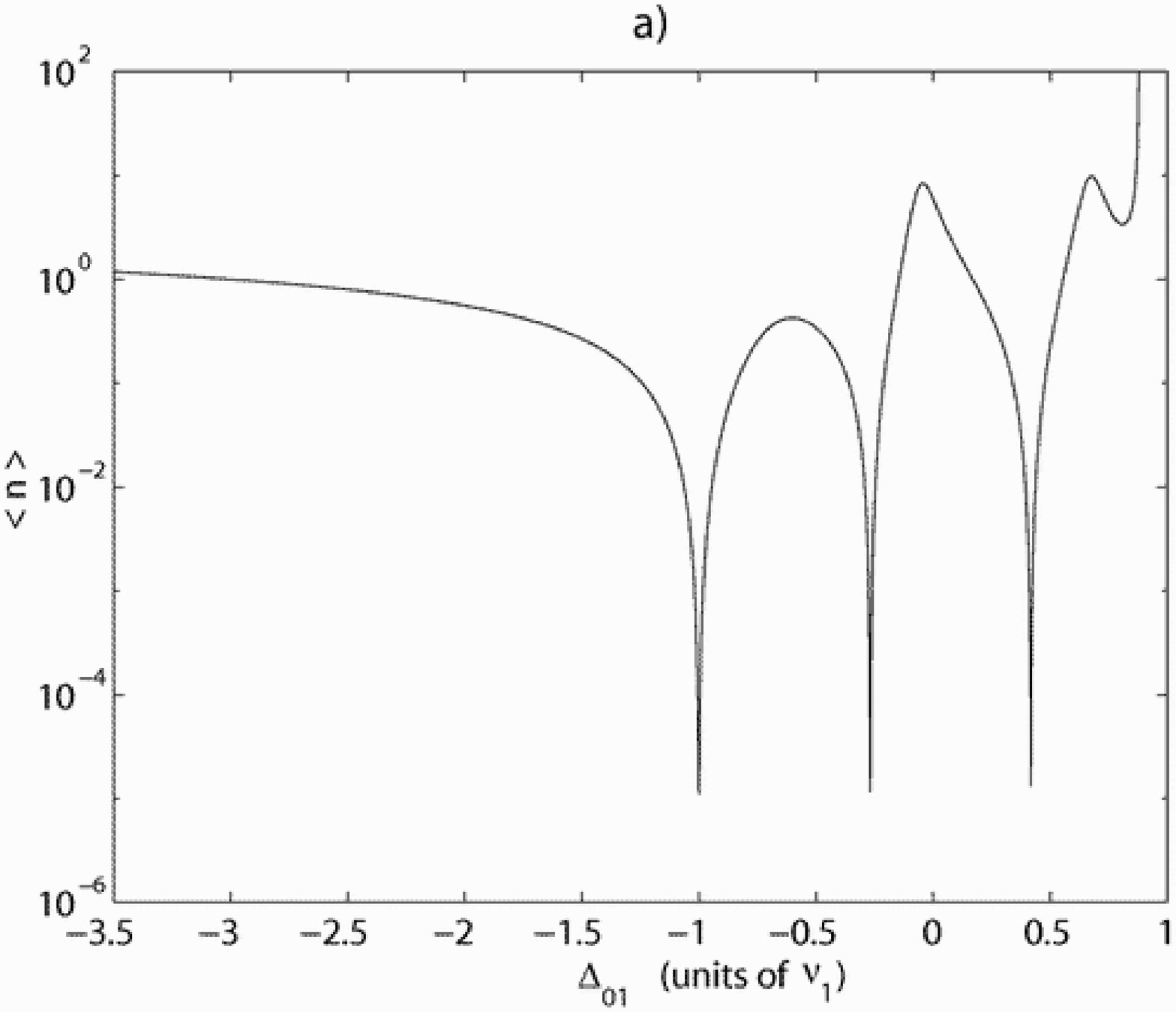}
\includegraphics[width=6cm]{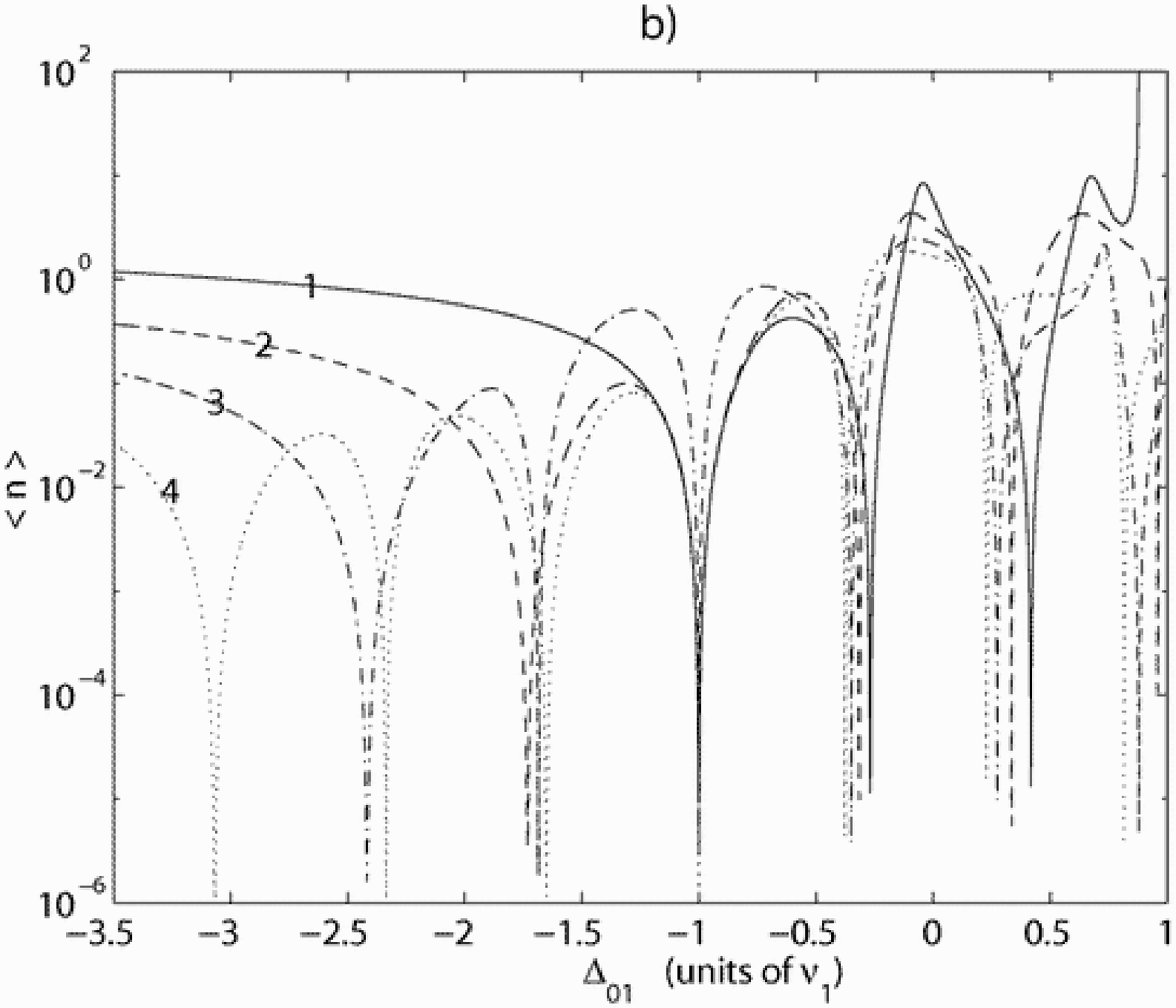}
\includegraphics[width=6cm]{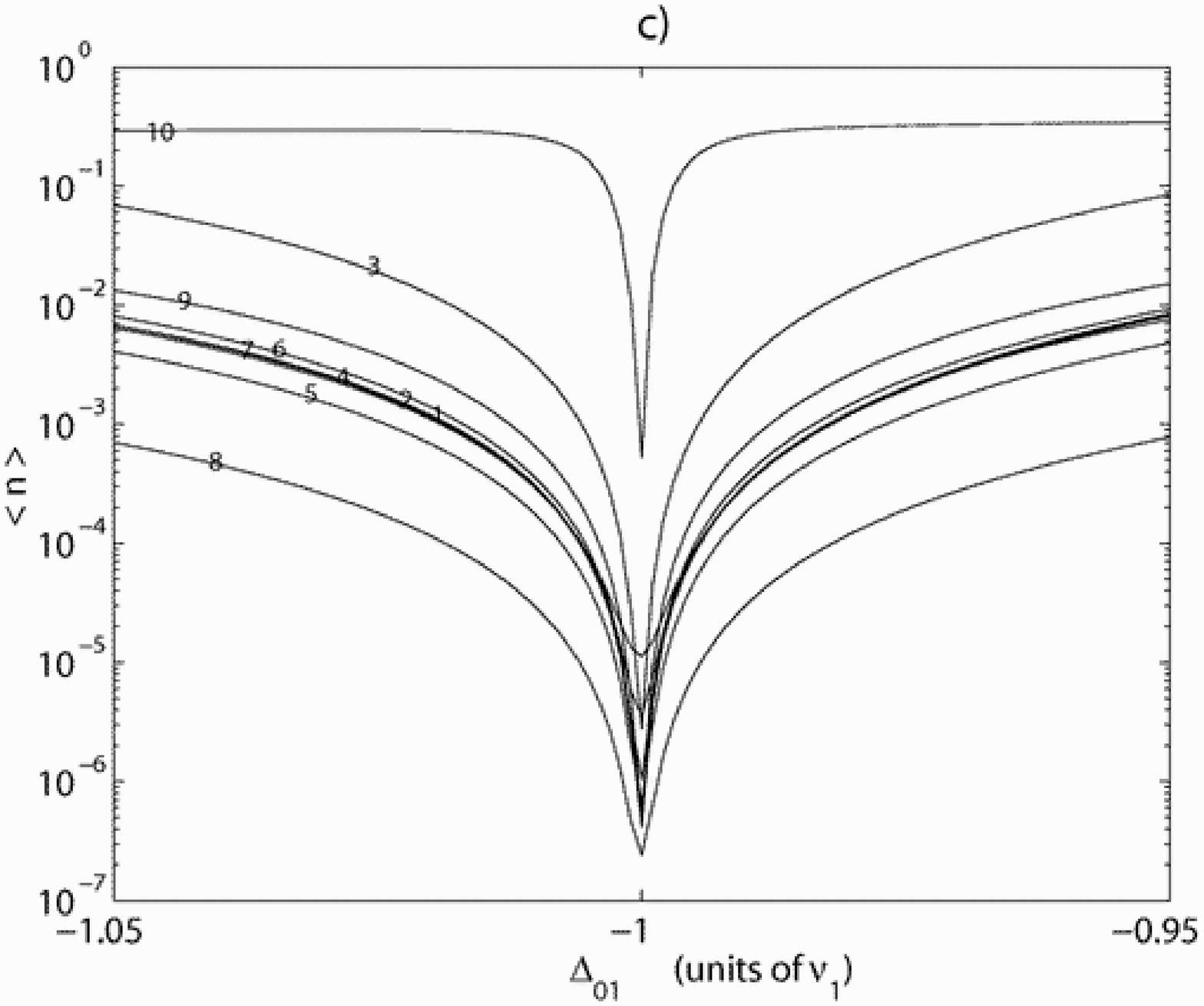}
\includegraphics[width=6cm]{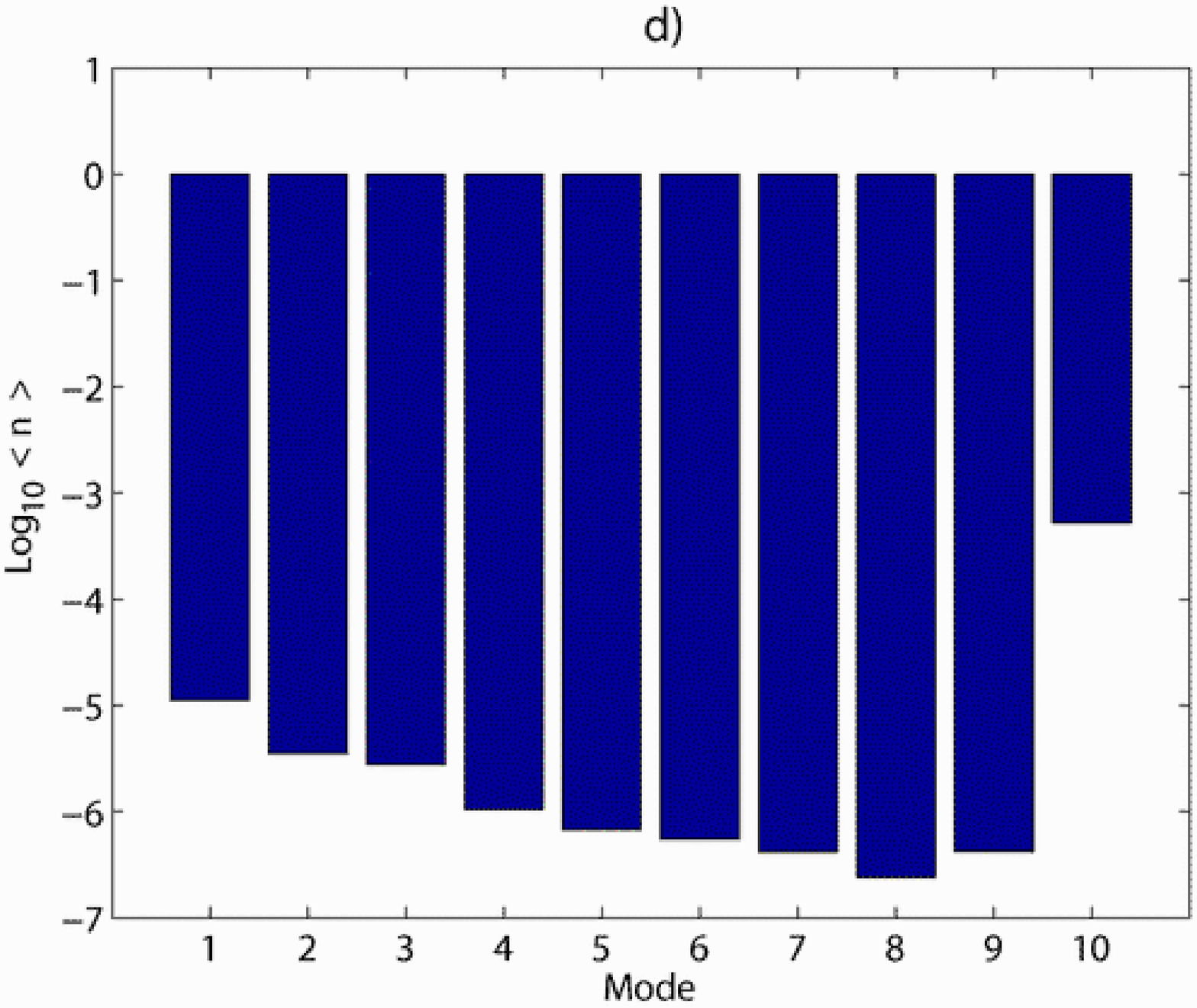}
\caption{Raman sideband cooling a chain of 10 ions in the presence
of a spatially inhomogeneous magnetic field such that the condition
$\omega_1-\nu_1=\ldots =\omega_N-\nu_N$ is fulfilled (see
Fig.~\ref{scheme}). The parameters are $\Omega_{12}=100\times
2\pi$kHz, $\Omega_{01}=5 \times 2\pi$kHz, $\Delta_{12} =-10\times
2\pi$MHz. The steady state mean vibrational excitation $\langle
n^{(\alpha)} \rangle$ is displayed. a) $\langle n^{(1)} \rangle$
(COM mode) as a function of $\Delta_{01}$ in units of $\nu_1$. b)
$\langle n^{(\alpha)} \rangle$ with $\alpha=1,2,3,4$ as  function of
$\Delta_{01}$. c) $\langle n^{(\alpha)} \rangle$ for
$\alpha=1,\ldots,10$ as a function of $\Delta_{01}$. d) $\langle
n^{(\alpha)}\rangle$ for $\alpha=1,\ldots,10$ when the chain is
simultaneously cooled at the detuning $\Delta_{01}=-\nu_1$.}
\label{cool_1}
\end{center}
\end{figure}

Figure~\ref{cool_1} displays the steady state mean vibrational
excitations that are obtained when the effective Rabi frequency for
the Raman coupling $\Omega_{01}=5 \times 2\pi$kHz, the Rabi
frequency of the repumper $\Omega_{12}=100\times 2\pi$kHz, and the
detuning $\Delta_{12}=-10\times 2\pi$MHz. Fig.~\ref{cool_1}a)
displays the mean vibrational quantum number $\langle n^{(1)}
\rangle$ of the COM mode as a function of the detuning
$\Delta_{01}$. Here three minima are visible. The leftmost minimum
occurs at $\Delta_{01}=-\nu_1$ and corresponds to resonance with the
red sideband of the COM in the spectrum of the first ion. The
minimum in the middle stems from the resonant drive of the red
COM-sideband in the spectrum of the second ion while the one on the
right is caused by the spectrum of the third ion in the chain. The
location of these resonances correspond to the ones shown in
Fig.~\ref{scheme}. Heating of the COM mode occurs if the blue
sideband of the COM mode is driven resonantly. In
Fig.~\ref{cool_1}a) the heating at the blue sideband of the first
ion, i.e.\ $\Delta_{01}=\nu_1$, is visible.

Figure~\ref{cool_1}b) displays $\langle n^{(1)}\rangle$, $\langle
n^{(2)}\rangle$, $\langle n^{(3)}\rangle$, and $\langle
n^{(4)}\rangle$ as a function of  $\Delta_{01}$. These mean
excitations have been calculated using the same parameters as in
Fig.~\ref{cool_1}a). The minima visible in this figure can be
identified with the corresponding resonances in the spectra of the
ions by comparison with Fig.~\ref{scheme}.  A common minimum occurs
at $\Delta_{01}=-\nu_1$ where all four vibrational modes are
simultaneously cooled to low excitation numbers.

The mean vibrational quantum number $\langle n^{(\alpha)} \rangle$
of all ten axial modes is displayed in Fig.~\ref{cool_1}c) as a
function of the detuning $\Delta_{01}$ in the neighborhood of the
value $\Delta_{01}=-\nu_1$. At this value of $\Delta_{01}$ the mean
excitation $\langle n^{(\alpha)} \rangle$ reaches its minimum for
all modes. The mode at frequency $\nu_{10}$ displays a relatively
narrow minimum and its mean vibrational number, although very small
at exact resonance, is orders of magnitude larger than the ones of
the other modes. In fact, ion 10 participates only little in the
vibrational motion of mode 10. This is described by the small matrix
element $S_{10}^{10}= 0.0018$ that scales the corresponding
Lamb-Dicke parameter as shown in Eq.~(\ref{eta}).

Fig.~\ref{cool_1}d) displays the steady state temperature of each
mode when the detuning of the Raman beams is set close to
$-\nu_1$. At this detuning the average excitation reaches its
minimum for each mode, which is $\langle
n^{(\alpha)}\rangle<10^{-3}$.

\subsubsection{Cooling rates for simultaneous Raman cooling}
\label{SimSeqCooling}

\begin{figure}[htbp]
\begin{center}
\includegraphics[width=6cm]{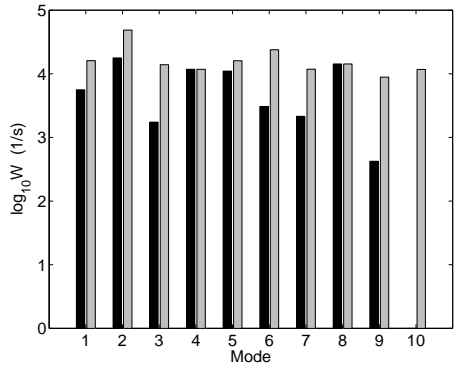}
\caption{Cooling rates for Raman sideband cooling a chain of 10
ions in the presence of a spatially inhomogeneous magnetic field
such that the condition $\omega_1-\nu_1=\ldots =\omega_N-\nu_N$ is
fulfilled. The same parameters have been used here as for
generating Fig. \protect\ref{cool_woB} ($\Omega_{12}=100\times
2\pi$kHz, $\Omega_{01}=30 \times 2\pi$kHz, $\Delta_{12} =-10\times
2\pi$MHz). The black bars indicate the rate for each mode when
{\em simultaneously} sideband cooling all modes. The grey bars
give the cooling rates that are achieved, if (still in the
presence of the magnetic field) the Raman detuning is set to that
value where the maximum cooling rate for each individual mode is
obtained.} \label{cool_wB}
\end{center}
\end{figure}

We now turn to the cooling rates that are
achieved when simultaneously cooling all modes, that is, the field
gradient leading to the spectrum in Fig.1 is applied, and
$\Delta_{01} =-\nu_1$. These rates are indicated by black bars in
Fig.~\ref{cool_wB}, and have been evaluated with the same parameters used
for the simulation of sequential cooling in Fig.~\ref{cool_woB}, that is,
$\Omega_{01}= 30\times 2\pi$kHz, $\Omega_{12}= 100\times 2\pi$kHz, and
$\Delta_{12}=-10\times 2\pi$MHz.

Figure~\ref{cool_wB} shows that the COM mode characterized by
$\nu_1$ and the mode characterized by $\nu_2$ are cooled most
efficiently, while modes 9 and 10 display much smaller cooling
rates. In particular, $\Gamma_{\rm cool}^{(10)}\approx
10^{-1}$s$^{-1}$ which will make cooling of this mode very slow at
$\Delta_{01} =-\nu_1$~\footnote{It is expected that mode 10 is much
less susceptible to heating by stray fields (the dominant heating
mechanism as discussed in section \ref{SeqCoolRate}) than the COM
mode \cite{Turchette00}. Therefore, such a low cooling rate will
suffice once this mode is close to its ground state. However, in
order to initially bring it close to the ground state the rate
$\Gamma_{\rm cool}^{10}$ needs to be larger.}.

The origin of this behaviour can be understood as follows. The
scheme of simultaneous sideband cooling requires that each ion of
the chain is employed to cool one of the axial modes. This is
achieved by applying a suitable magnetic field gradient. The
simplest experimental implementation uses a monotonically increasing
magnetic field, such that the first ion of the chain is used for
cooling mode 1, the second ion mode 2, etc. (compare Fig. 1).
However, the coupling of a certain ion displacement to a certain
mode can be very small. It occurs, for instance, that the largest
axial excitations couple weakly to the ions at the edges of the ion
string, instead they are mainly characterized by oscillations of
ions in the center of the chain~\cite{PRL04}. A manifestation of
this behavior is the small value of the matrix element
$S_{10}^{10}$. Thus, mode 10 is not efficiently cooled by
illuminating ion 10, rather it is optimally cooled by addressing an
ion closer to the center of the chain. This is evident by inspection
of the grey bars in Fig.~\ref{cool_wB} that indicate the {\em
optimal} cooling rate for each individual mode, obtained by
employing that particular ion in the chain which has the largest
coupling to the mode to be cooled. In presence of the magnetic field
gradient, this optimal cooling rate is achieved, if the detuning,
$\Delta_{01}$ is set such that the appropriate red-sideband
transition of this particular ion is driven resonantly. In this way,
the cooling rate is maximal for that particular mode (i.e., each
grey bar corresponds to a different detuning).

For the case of mode 10 the difference between the cooling rates
depending on which ion is addressed is particularly striking: the
cooling rate of mode 10 at detuning $\Delta_{01}= -\nu_1$ (black
bar) is very small, as noted above, whereas the optimal rate
$\Gamma_{\rm cool}^{(10)}= 2.63\times 10^{3}$s$^{-1}$  (given the
parameters used here) is achieved at $\Delta_{01}= -3.28\times
\nu_1$. At this detuning ion 6 is used for cooling mode 10, as can
be seen from Fig.~\ref{scheme}.

Efficient cooling of all modes, given the field gradient that gives
rise to the spectrum illustrated in Fig.~1, can be obtained by
combining sequential and simultaneous cooling as follows: First the
ion string is illuminated with radiation such that
$\Delta_{01}=-3.28\times \nu_1$ which will efficiently cool mode 10
and will have little effect on all other modes. Then, we apply
radiation with $\Delta_{01}=-\nu_1$ (cooling rates indicated by
black bars in Fig. ~\ref{cool_wB}). This will simultaneously cool
all other modes. This is a specific recipe to cool 10 ions. For an
arbitrary number of ions, first one cools the modes which cannot be
efficiently simultaneously cooled with all other modes, and then, in
a second step, as many modes as possible are cooled simultaneously
(in our case modes 1 through 9).

A necessary condition for the efficiency of this scheme is
\begin{equation}
\Gamma_{\rm heat}^{(\alpha)} \ll \Gamma_{\rm cool,sim}^{(\alpha)} \
\forall \ \alpha=1,\ldots ,10 \ .
 \label{CoolHeat_sim}
\end{equation}
where $\Gamma_{\rm cool,sim}^{(\alpha)}$ is the cooling rate  of
mode $\alpha$ when simultaneously cooling it with the other modes.
Condition~(\ref{CoolHeat_sim}) is the analogue to~(\ref{CoolHeat})
which we derived for sequential cooling. Moreover, since in the
scheme proposed here modes 1 through 9 are simultaneously cooled, a
condition analogous to the one in Eq.~(\ref{SeqCool}) has to be
fulfilled {\em for mode 10 only}: After mode 10 has been cooled,
this mode may not heat up again appreciably during the time for
which modes 1 through 9 are simultaneously cooled. This condition is
expressed as
\begin{equation}
\Gamma_{\rm heat}^{(10)} \times T_{\rm sim}^{(10)} \ll 1 \ .
\label{SimCool}
\end{equation}
where $T_{\rm sim}^{(10)}$ is the time during which modes 1 through
9 are cooled simultaneously to the desired final values, that is,
the time during which mode 10 is {\em not} cooled and could
heat up. Here, it is given by
\begin{equation}
T_{\rm sim}^{(10)}= {\rm max_{\beta=1\ldots 9}}\left[\ln
\frac{\langle n^{(\beta)}\rangle_i} {\langle n^{(\beta)}\rangle_f} /
(\Gamma_{\rm cool}^{(\beta)}-\Gamma_{\rm heat}^{(\beta)})\right] \
,\label{T_heat_sim}
\end{equation}
This relation is analogous to relation~(\ref{T_heat_seq}) for
sequential cooling.

The times $T_{\rm seq}^{(10)}$ (eq.~(\ref{T_heat_seq})) and $T_{\rm
sim}^{(10)}$ (eq.~(\ref{T_heat_sim})) give an upper limit for the
tolerable trap heating rate of mode 10 for the cases of sequential
and simultaneous cooling, respectively. It turns out that the
relevant time scales $T_{\rm seq}^{(10)}$ and $T_{\rm sim}^{(10)}$
are of the same order of magnitude: Using eq. \ref{T_heat_sim} to
compute $T_{\rm sim}^{(10)}$ for the parameters used in Fig.
\ref{cool_wB} gives $T_{\rm sim}^{(10)}\approx \ln(\langle n
\rangle_i/\langle n \rangle_f) \times 2.4\times 10^{-3}$s, while
$T_{\rm seq}^{(10)}\approx \ln(\langle n \rangle_i/\langle n
\rangle_f) \times 1.3\times 10^{-3}$s is obtained from
eq.~(\ref{T_heat_seq}) for the same set of
parameters~\cite{footnoteC}. Thus, the cooling scheme proposed here
does not increase the admissible trap heating rate of mode 10. Since
this rate is expected to be low anyway, this will not be a
restriction prohibiting the successful cooling of a long ion chain
using {\em either} sequential or simultaneous cooling.

\subsection{Comparison of sequential and simultaneous cooling}
\label{Sec:Compare}

So far we have stated the general conditions that have to be met  in
order to efficiently cool an ion chain with sequential cooling and
with a scheme combining simultaneous and sequential sideband
cooling. In this section we discuss their efficiencies.

We note that when using sequential sideband cooling, one may utilize
all ions in the chain in order to cool one mode, where the cooling
rates of each ion add up incoherently. In the case of simultaneous
sideband cooling, on the other hand, only one ion is employed in
order to cool a particular mode. Assuming that the coupling of this
ion to the mode is sufficiently large to allow for efficient
cooling, the following expression for the average simultaneous
cooling rate is deduced
\begin{equation}
\Gamma_{\rm cool,sim} \approx \frac{1}{N}\Gamma_{\rm cool,seq} \
 \label{G_simseq}
\end{equation}
where $\Gamma_{\rm cool,seq}$ is the cooling rate for sequential
cooling averaged over all modes. Thus, for simultaneous cooling
relation~(\ref{CoolHeat_sim}) yields
\begin{equation}
\Gamma_{\rm heat}^{(\alpha)} \ll \frac{1}{N}\Gamma_{\rm cool,seq}
\equiv \Gamma_{\rm <,sim}\ ,
 \label{CoolHeat_sim2}
\end{equation}
From~(\ref{G_simseq}), since the total cooling rate is
$\Gamma_{\rm <,seq}= \Gamma_{\rm cool,seq}/N$, one obtains that
the efficiencies of simultaneous and sequential cooling are
comparable. However, it should be remarked that this estimate
corresponds to the worst case for simultaneous cooling. In fact,
estimate~(\ref{G_simseq}) is correct for long-wavelength
vibrational excitations, which correspond to low-frequency axial
modes, where practically all ions of the chain participate in the
mode oscillation. Short-wavelength excitations, on the other hand,
are characterized by large displacements of the central ions,
while the ions at the edges practically do not move~\cite{PRL04}.
Due to this property, for these modes one may find a magnetic
field configuration such that $\Gamma_{\rm cool,sim}^{(\alpha)}
\sim \Gamma_{\rm cool,seq}^{(\alpha)}$.

On the basis of these qualitative considerations, one may in
general state that simultaneous cooling of the chain is at least
as efficient as sequential cooling. For the configuration
discussed in this paper the two methods are comparable.
Differently from sequential cooling, however, the efficiency of
simultaneous cooling can be substantially improved by choosing a
suitable magnetic field configuration that maximizes coupling of 
each mode to one ion of the chain.

\section{Sideband cooling using microwave fields}
\label{MW}

\subsection{Effective Lamb-Dicke parameter}

We investigate now sideband cooling of the ion
chain's collective motion using microwave radiation for driving
the sideband transition. It should be remarked that in this
frequency range sideband excitation cannot be achieved by means
of photon recoil which for long wavelengths is negligible. This is
evident from eq. \ref{eta} using a typical trap frequency,
$\nu_\alpha$ of the order $2\pi\times 1$MHz. Nevertheless, in
\cite{Mintert01} it was shown that with the application of a
magnetic field gradient an additional mechanical effect can be
produced accompanying the absorption/emission of a photon. This is
achieved by realizing different mechanical potentials for the
states $|1\rangle$ and
$|0\rangle$~\cite{Mintert01,Wunderlich02,Wunderlich03}. Hence, by
changing an ion's internal state by stimulated absorption or
emission of a microwave photon the ion experiences a mechanical
force.

An effective LDP can be associated with this force, that is defined
as~\cite{Wunderlich02}:
\begin{equation}
\eta_{j\alpha}^{\rm eff}\ e^{i\varphi_j} \equiv \eta_j^\alpha +
i\varepsilon_j^\alpha
\label{etaEff}
\end{equation}
where
\begin{equation}
\label{eta_mu} \varepsilon_{j}^{\alpha}= S_{j}^\alpha \frac{
\partial_z \omega_j\,\Delta\! z_\alpha}{\nu_\alpha} \ ,
\end{equation}
Here $\partial_z\omega_j$ is the spatial derivative of the
resonance frequency of the ion at $z_j^{(0)}$ with respect to
$z_j$, and $\Delta\! z_\alpha = \sqrt{\hbar/(2m\nu_\alpha)}$. The reader is
referred to the appendix for the theoretical description of the ion chain
dynamics in presence of a spatially-varying magnetic field.

The term $\eta_j^{\alpha}$ appearing in Eq.~(\ref{etaEff}) is the LDP due to
photon recoil and defined in Eq.~(\ref{eta}), while the term
$\varepsilon_j^\alpha$ is the LPD arising from the mechanical effect induced
by the magnetic field gradient. Their ratio is given by
\begin{equation}
\label{ratio}
\frac{\varepsilon_j^{\alpha}}{\eta_j^{\alpha}}=
\frac{\kappa_j^{\alpha}}{k \Delta\! z_\alpha} =
\frac{1}{2\pi}\frac{\lambda}{\Delta\! z_\alpha}
\kappa_j^{\alpha}
\end{equation}
where $\lambda$ is the wavelength of the considered transition and
$\kappa_j^{\alpha}=\Delta\! z_\alpha \partial_z \omega_j/\nu_{\alpha}$
is the rescaled frequency gradient \cite{Mintert01,Wunderlich02}.

For a transition in the microwave frequency range, like the
transition between the states $|0\rangle$ and $|1\rangle$ in \Yb,
using typical values, like $\kappa_j^{\alpha} \approx 10^{-3}$,
$\lambda\approx 10^{-2}$m, $\Delta\! z_\alpha \approx 10^{-8}$m,
one finds that the Lamb-Dicke parameter due to the recoil of a
microwave photon is at least two orders of magnitude smaller than
the one due to the mechanical effect induced by the magnetic
field, and thus $\eta_{j\alpha}^{\rm eff} \approx
\varepsilon_j^{\alpha}$ in the microwave region.

\subsection{Steady state population and cooling rates}
\label{SimCoolingMW}

We study sideband cooling using microwave radiation using the model
outlined in the appendix. For the Rabi frequencies the same
parameters as in section \ref{Optical} are employed (i.e.,
$\Omega_{01}=5 \times 2\pi$kHz, $\Omega_{12}=100\times 2\pi$kHz,
$\Delta_{12}=-10\times 2\pi$MHz). The effective LDP is determined by
the magnitude of the magnetic field gradient that is used to
superimpose the motional sidebands of all vibrational modes, and is
given by $\eta_{j\alpha}^{\rm eff} \approx \varepsilon_j^{\alpha}=
S_j^\alpha \kappa_j^\alpha \approx S_j^\alpha \times 2\times
10^{-3}$.

It must be remarked that these parameters are outside the range of
validity for the application of the numerical method we use. In
fact, by applying it one neglects the fourth and sixth orders in the
LDP expansion of the optical transition of the repumping cycle,
which are of the same order of magnitude as the microwave sideband
excitation. Nevertheless, given the complexity of the problem,
characterized by a large number of degrees of freedom, we have
chosen to use this simpler method in order to get an indicative
estimate of the efficiency as compared to the case in which the
sidebands are driven by optical radiation. Therefore, the results we
obtain in this section are indicative. In fact, neglecting higher
orders in the LDP of the optical transition corresponds to
underestimate heating effects due to diffusion.

\begin{figure}[htbp]
\begin{center}
\includegraphics[width=6cm]{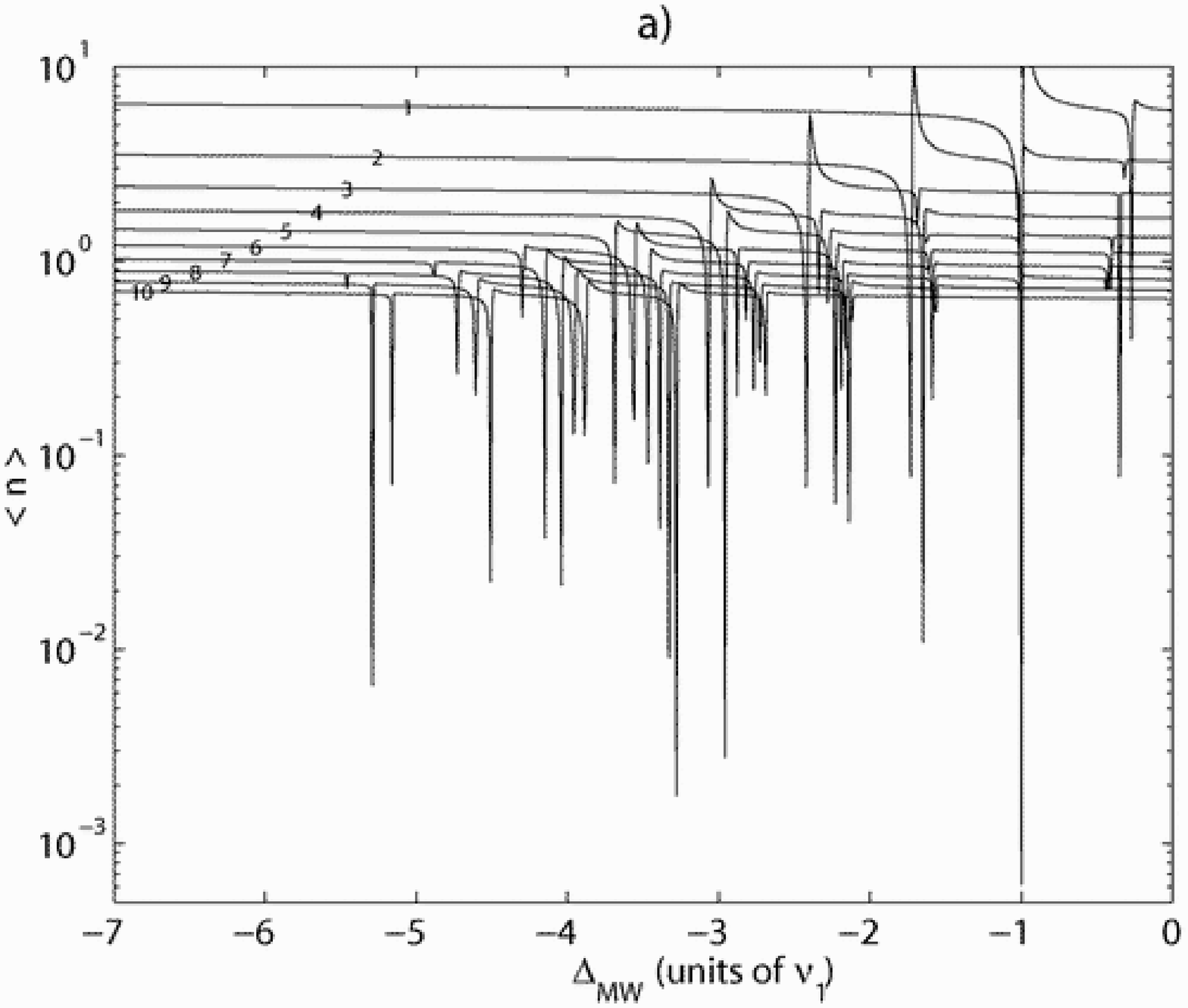}
\includegraphics[width=6cm]{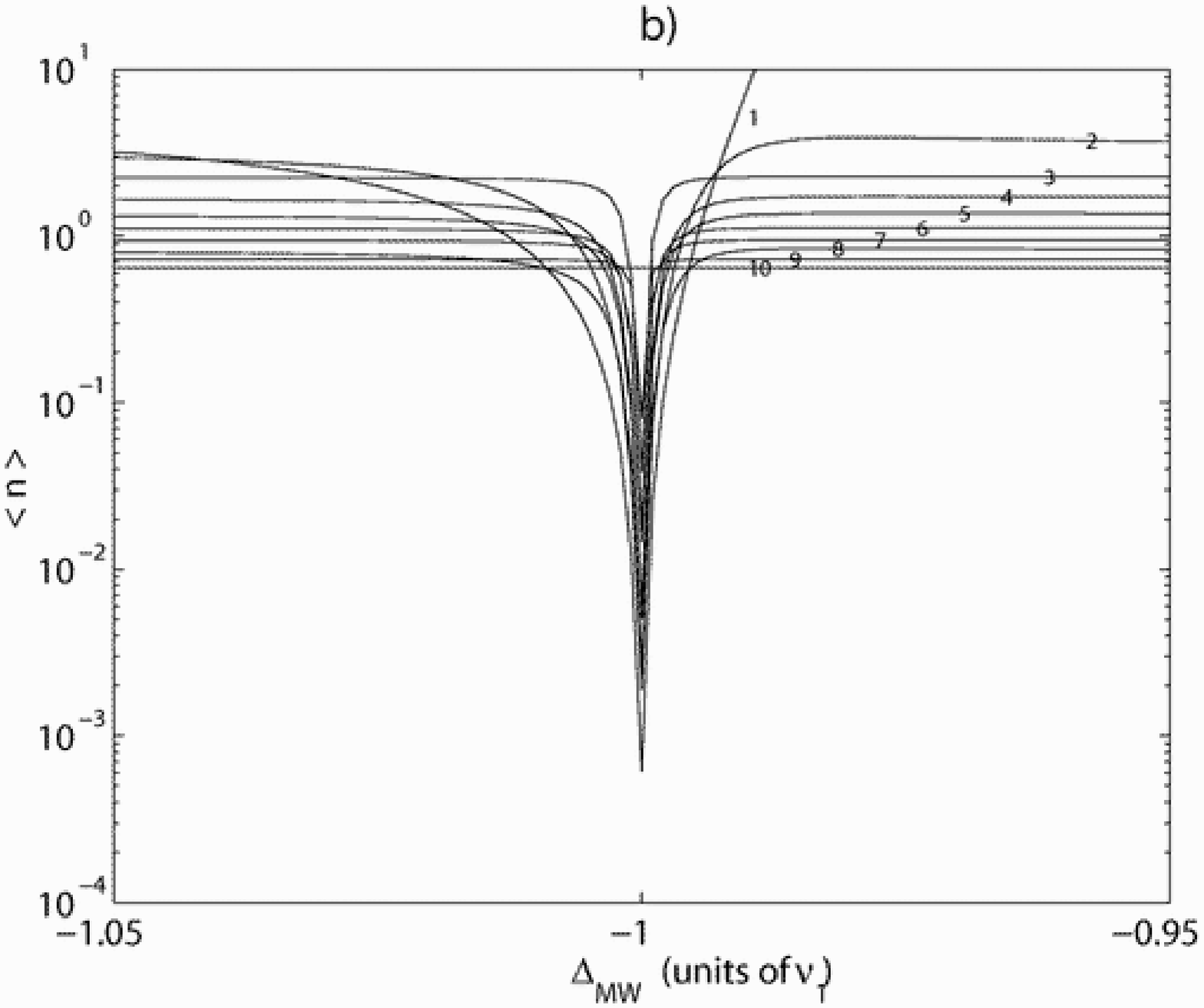}
\includegraphics[width=6cm]{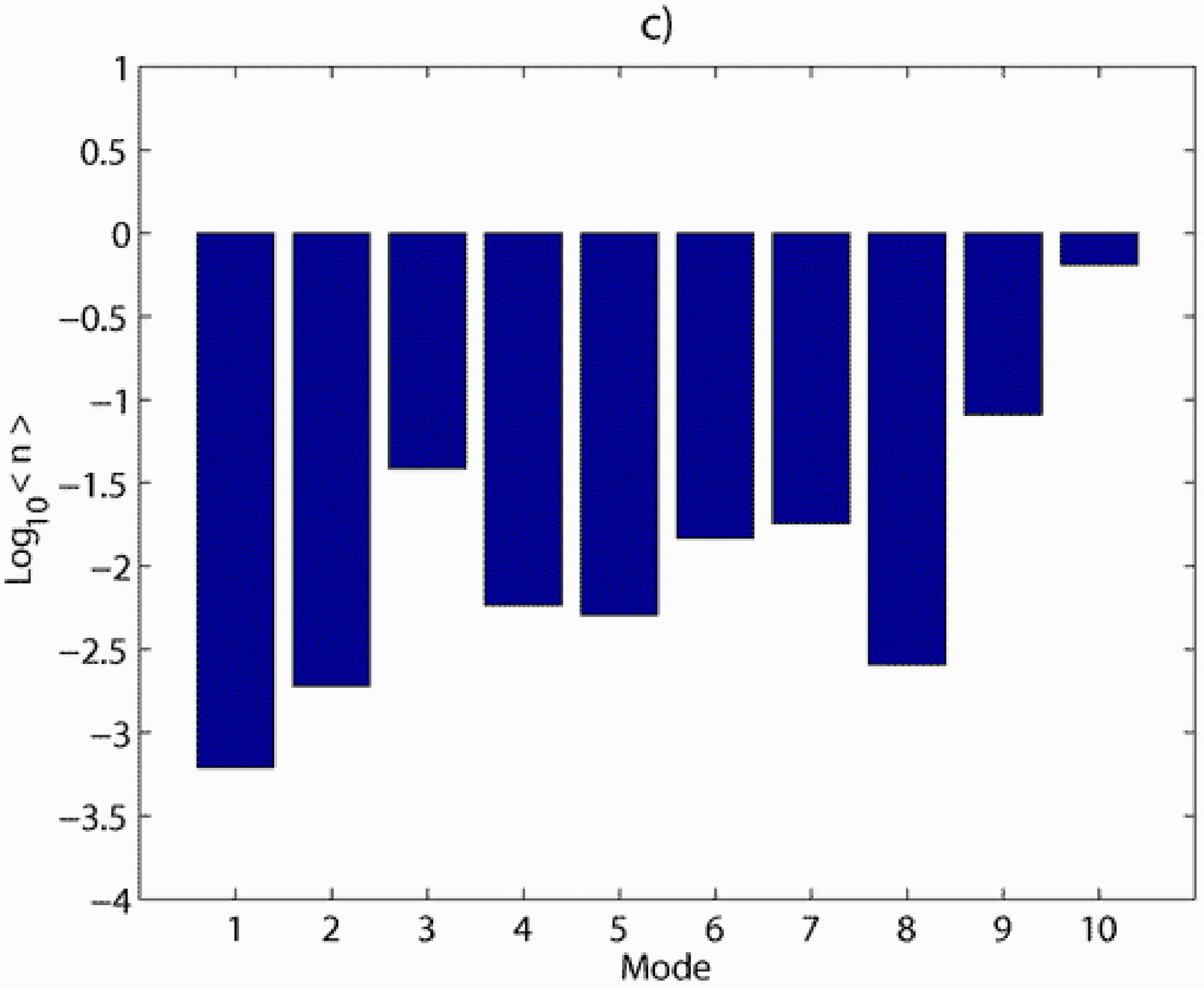}
\includegraphics[width=6cm]{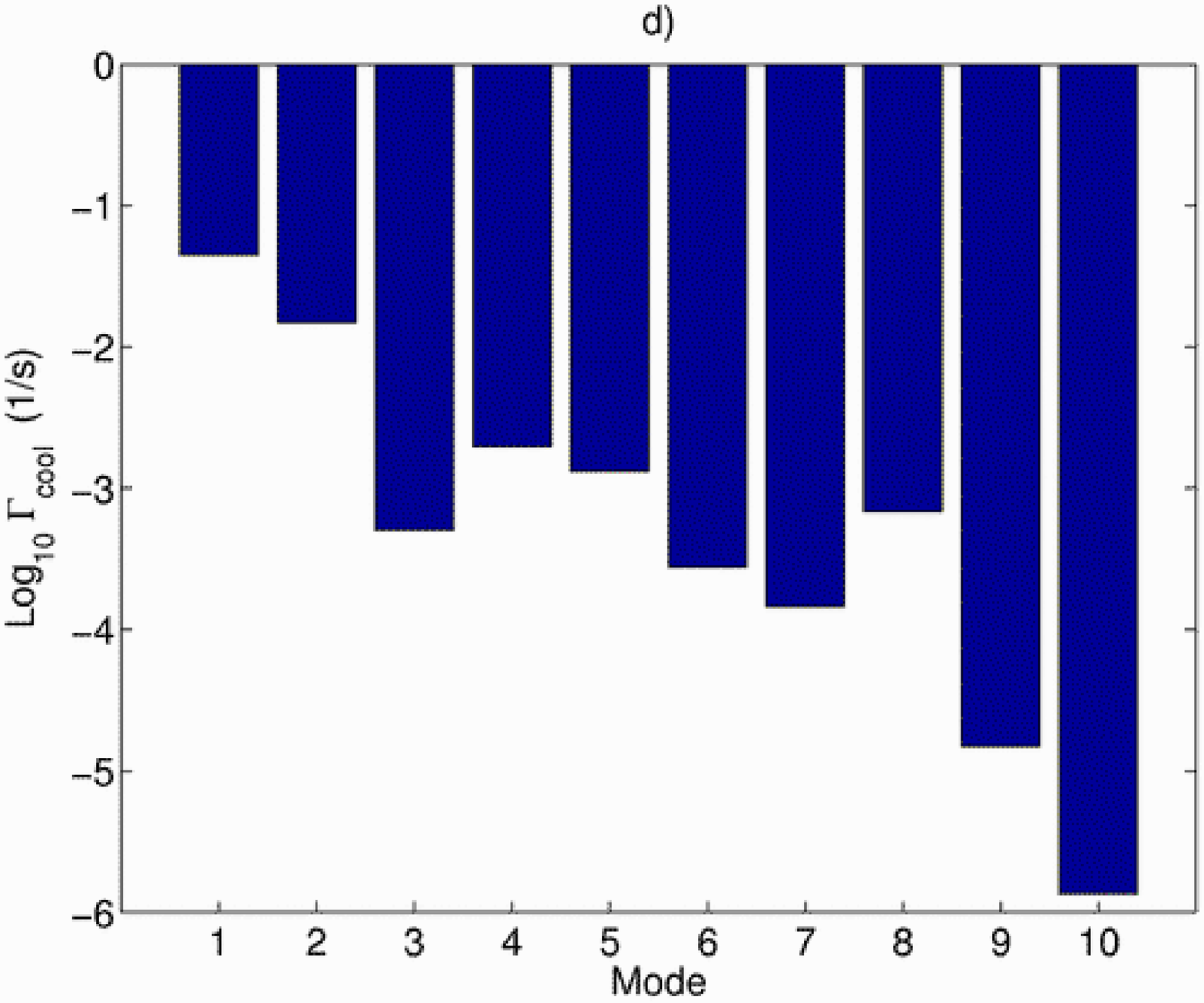}
\caption{Microwave sideband cooling a chain of 10 ions in the
presence of a spatially inhomogeneous magnetic field such that the
condition $\omega_1-\nu_1=\ldots =\omega_N-\nu_N$ is fulfilled, that
is, {\em simultaneous} sideband cooling using microwave radiation
for sideband  excitation. Same parameters as for Fig.
\protect\ref{cool_1} ($\Omega_{12}=100\times 2\pi$kHz,
$\Omega_{01}=5 \times
  2\pi$kHz, $\Delta_{12} =-10\times 2\pi$MHz). a) Mean vibrational numbers
$\langle n^{(\alpha)} \rangle_f$ at steady state as a function of
the detuning $\Delta_{01}$. b) Same as a) for the values around
$\Delta_{01}=-\nu_1$. c) $\langle n^{(\alpha)} \rangle_f$ at
$\Delta_{01}=-\nu_1$, i.e.\ when all ten vibrational modes are
simultaneously cooled. d) Cooling rates for $\Delta_{01}=-\nu_1$.}
\label{cool_2}
\end{center}
\end{figure}

Fig.~\ref{cool_2}a) shows the steady state mean vibrational quantum
number $\langle n^{(\alpha)}\rangle_f$ of the 10 axial vibrational
modes as a function of the detuning $\Delta_{01}$ of the microwave
radiation, where $\Delta_{01}=0$ when the microwave field is at
frequency $\omega_1$. Figure~\ref{cool_2}b) displays the final
excitation number of all vibrational modes as a function of
$\Delta_{01}$ around the value $\Delta_{01}= -\nu_1$.
Figure~\ref{cool_2}c) shows the steady state mean excitation of all
10 modes at $\Delta_{\mathrm MW}= -\nu_1$.

All vibrational modes are cooled close to their ground state.
However, the average occupation number of the highest vibrational
frequency $\nu_{10}$ is orders of magnitude larger than the ones of
the other modes. The origin of this behavior is the small value of
the coefficient $S_{10}^{10}= 0.0018$ as is discussed in
Sec.~\ref{Optical}. A comparison of Fig. \ref{cool_2}c) with the
results obtained by simultaneously cooling using an optical Raman
processes (Fig.~\ref{cool_1}d) shows that the mean vibrational
numbers that are achieved here are considerably larger.

In Fig. \ref{cool_2}d) the cooling rates are displayed that are
obtained when driving the sideband transition with microwave
radiation. These rates are much smaller than the ones shown in
Fig.~\ref{cool_wB} that are obtained using an optical Raman process.

From this comparison it is evident that simultaneous cooling is more effective
by using an optical Raman process than microwave radiation. In particular, the cooling
rates obtained with simultaneous microwave sideband cooling can be
comparable to the heating rates in some experimental situations,
resulting in inefficient cooling.

This striking difference in the efficiency can already be found, if
comparing the two methods when they are applied to cooling a single
ion. Its origin lies in the fact that in the optical case the LDP
accounting for the photon recoil, namely $\eta_j^{\alpha}$ in
Eq.~(\ref{etaEff}), is considerably larger than the LDP,
$\varepsilon_j^\alpha$ caused by the magnetic field gradient used
here to superimpose the motional sidebands. This can be verified by
using an optical wavelength in Eq.~(\ref{ratio}) which gives
$\eta_{j\alpha}^{\rm eff}\approx \eta_j^{\alpha}$. On the other
hand, driving the $|0\rangle-|1\rangle$ transition directly by
microwaves results in $\eta_{j\alpha}^{\rm eff} \approx
\varepsilon_j^{\alpha}\ll \eta_j^{\alpha}$. For the model system
considered here, the optical LDP is at least one order of magnitude
larger than the microwave LDP for simultaneous cooling.

In the microwave sideband cooling scheme optical transitions are
used for repumping, thus leading to an enhanced diffusion rate
during the dynamics and thus to lower efficiencies. The fundamental
features of these dynamics can be illustrated by a rate equation of
the form Eq.~(\ref{Palpha}), describing sideband cooling of a single
ion with microwave radiation, where the rates are
\begin{eqnarray}
  \label{Yb:1} &&A_{-} =\frac{\Omega^2}{2\gamma} \Bigl[|\varepsilon|^2+\phi
  |\eta_{\rm opt}|^2\frac{\gamma^2}{4\nu^2+\gamma^2} \Bigr]\\ &&A_{+}
  =\frac{\Omega^2}{2\gamma}
  \Bigl[|\varepsilon|^2\frac{\gamma^2}{16\nu^2+\gamma^2}+\phi |\eta_{\rm
    opt}|^2\frac{\gamma^2}{4\nu^2+\gamma^2} \Bigr]\ , \label{Yb:2}
\end{eqnarray}
$\eta_{\rm opt}$ accounts for the recoil due to the spontaneous
emission when the ion is optically pumped to the state
$\0$~\cite{Morigi01}, and $\varepsilon$ is the LDP for the microwave
transition due to the field gradient. The latter multiplies the
terms where a sideband transition occurs by microwave excitation,
whereas $\eta_{\rm opt}$ multiplies the terms where sideband
excitation occurs by means of spontaneous emission, which thus
describe the diffusion during the cooling process. Efficient ground
state cooling is achieved when the rate of cooling is much larger
than the rate of heating, which corresponds to the condition $A_-/
A_+ \gg 1$. For $\nu_1\gg\gamma_{\mu}$, whereby $\gamma_{\mu}$ is
the linewidth of the $|0\rangle\to|1\rangle$ transition, this ratio
scales as
\begin{equation}
  \frac{A_-}{A_+}\sim \frac{|\varepsilon|^2}{|\eta_{\rm
      opt}|^2}\frac{4\nu^2}{\gamma^2}
\end{equation}
This result differs for the ratio obtained in the all-optical case,
where $A_-/ A_+ \sim 4\nu^2/\gamma^2$. For typical parameters,
corresponding to a magnetic field gradient that superposes all
sidebands in an ion trap, $|\varepsilon| \ll |\eta_{\rm opt}|$.
Hence, for a given value of the ratio $\gamma/\nu$ the cooling
efficiency in the optical case is considerably larger than in the
microwave case.

Note that the parameter $\varepsilon$ can be made larger by
increasing the magnitude of the magnetic field gradient, which in
Eq.~(\ref{ratio}) corresponds to increasing $\kappa_j^\alpha$.
However, if the modes of an ion chain are to be cooled
simultaneously,  the choice of the magnetic field gradient is fixed
by the distance between neighboring ions, and the efficiency is thus
limited by this requirement.

If an individual ion (or a neutral atom confined, for example, in an
optical dipole trap) is to be sideband cooled, or sequential cooling
is applied to a chain of ions, then the above mentioned restriction
on the magnitude of $\varepsilon$ is not present and microwave
sideband cooling can be as efficient as Raman cooling. For this case, 
method~\cite{Marzoli94} may be implemented, provided the Lamb-Dicke regime applies
and $\eta_{\rm opt}$ and $\epsilon$ are of the same order of magnitude.

\section{Experimental considerations}
\label{Exp}

In this section we discuss how the magnetic field gradient for
simultaneous sideband cooling can be generated and how cooling is
affected, if the red motional sidebands of different vibrational
modes are not perfectly superposed. In order to demonstrate the
feasibility of the proposed scheme it is sufficient to restrict the
discussion to very simple arrangements of magnetic field generating
coils.

The use of a position dependent ac-Stark shift has been proposed in
\cite{Staanum02} to modify the spectrum of a linear ion chain. This
may be another way of appropriately shifting the sideband resonances
but will not be considered here.

\subsection{Required magnetic field gradient}

If the vibrational resonances and the ions were equally spaced in
frequency and position space, respectively, then a constant field
gradient, appropriately chosen, could make all $N$ modes overlap
and let them be cooled at the same time. Since $\nu_\alpha -
\nu_{\alpha-1}$ decreases monotonically with growing $\alpha$ and
the ions' mutual distances vary with $z_j$, the magnetic field
gradient has to be adjusted along the $z-$axis. The field gradient
needed to shift the ions' resonances by the desired amount is
obtained from
 \begin{eqnarray}\label{bz}
  \left.\frac{\partial B}{\partial z}\right|_{(z_j+z_{j-1})/2}
&  \approx &\frac{B(z_j)-B(z_{j-1})}{z_j-z_{j-1}} \nonumber \\
  &  \stackrel{!}{=}& \frac{\upsilon_j-\upsilon_{j-1}}{\zeta_j-\zeta_{j-1}}
                        \zeta_0\nu_1 \frac{\hbar}{\mu_B},\
                        j=2,\ldots,N
 \end{eqnarray}
where $\zeta_j\equiv z_j/\zeta_0$ is the scaled equilibrium
position of ion $j$, and $\upsilon_j$ is the square root of the
$j-$th eigenvalue of the dynamical matrix. Eq. \ref{bz} describes
the situation for moderate magnetic fields (the Zeeman energy is
much smaller than the hyperfine splitting), such that
 $\partial_z\omega_j = 1/2 \;g_J\mu_B\partial_z B$ with $g_J\approx g_s=2$
(state $\0$ does not depend on $B$). As an example, we consider
again a string of $N=10$ \Yb ions in a trap characterized by
$\nu_z =1 \times 2\pi $MHz (thus, $\zeta_0 = 2.7\mu$m).

\begin{figure}[htbp]
\begin{center}
\includegraphics[width=6cm]{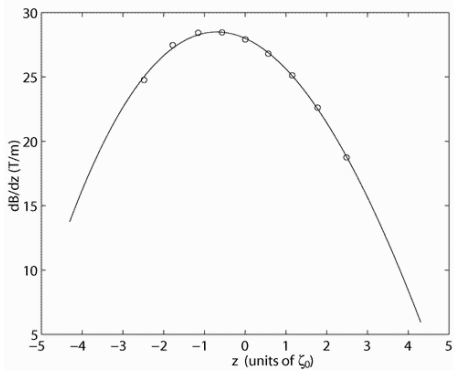}
\caption{Required magnetic field gradient to superpose the
motional red sidebands of ten $^{171}$Yb$^+$ ions (markers) in a
trap characterized by a COM frequency $\nu_1 = 1\times 2\pi $MHz
and calculated field gradient (solid line) produced by three
single wire windings (see text).} \label{FigGradient}
\end{center}
\end{figure}

The markers in Fig. \ref{FigGradient} indicate the values of the
required field gradient according to eq. \ref{bz} whereas the solid
line shows the gradient generated by 3 single windings of diameter
100$\mu$m, located at $z=-100, 50, $ and $100$ \umu m $\approx
36\zeta_0$, respectively (the trap center is chosen as the origin of
the coordinate system) \cite{Fortagh98}. Running the currents
-5.33A, -6.46A, and 4.29A,respectively, through these coils produces
the desired field gradient at the location of the ions. Micro
electromagnets with dimensions of a few tens of micrometers and
smaller are now routinely used in experiments where neutral atoms
are trapped and manipulated \cite{Drndic98}. Current densities up to
$10^8$A/cm$^2$ have been achieved in such experiments. A current
density more than two orders of magnitude less than was achieved in
atom trapping experiments would suffice in the above mentioned
example.

This configuration of magnetic field coils shall serve as an example
to illustrate the feasibility of the proposed cooling scheme in what
follows. It will be shown that with such few current carrying
elements in this simple arrangement one may obtain good results when
simultaneously sideband cooling all axial modes. More sophisticated
structures for generating the magnetic field gradients can of course
be employed, making use of more coils, different diameters, variable
currents, or completely different configurations of current carrying
structures.

\begin{figure}[htbp]
\begin{center}
\includegraphics[width=7.5cm]{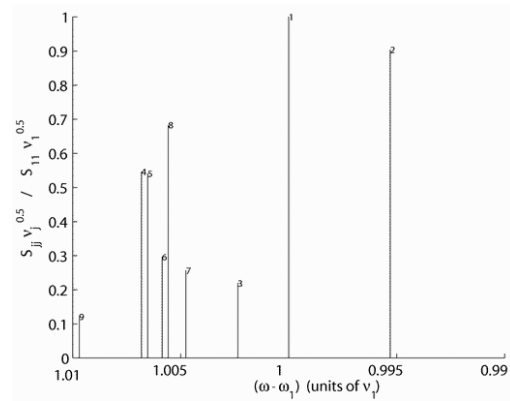}
\caption{The vertical bars indicate the frequencies $\omega_j
-\nu_j$ of first order sideband resonances corresponding to 10 axial
vibrational modes of 10 $^{171}$Yb$^+$ ions in an ion trap with the
field gradient shown in Fig. \ref{FigGradient}. The frequencies are
given relative to the transition frequency $\omega_1$ of the first
ion, and the bars are labelled with $j=1,\ldots , 10$. The height of
the bars indicates the transition probability associated with a
given sideband relative to the first sideband (COM mode) of ion 1
for a Raman transition. The relative transition probability is
proportional to $|\eta_{j\alpha}^{\rm eff}|/|\eta_{11}^{\rm eff}|
\approx S_j^\alpha \nu_\alpha^{-(1/2)}/ S_1^1 \nu_1^{-(1/2)}$ with
$j=\alpha$} \label{FigSideband}
\end{center}
\end{figure}

Ideally, all 10 sideband resonances would be superimposed for
optimal cooling. The resonances shown in Fig. \ref{FigSideband}
result from the field gradient calculated using the simple field
generating configuration described above, and do not all fall on
top of each other. Nevertheless, Fig. \ref{FigSideband} shows how
well all 10 sideband resonances are grouped around $\omega_1 -
\nu_1$. Vertical bars indicate the location relative to $\omega_1$
of the red sideband resonance, $\omega_j-\nu_j$ of the $j-$th ion,
with $j=1,\ldots ,10$. These resonances all lie within a frequency
interval of about $0.015 \times \nu_1 /2\pi = 15$kHz.

The height of the bars in Fig. \ref{FigSideband} indicates the
strength of the coupling between the driving radiation and the
respective sideband transition relative to the COM sideband of ion
number 1. The relevant coupling parameter is the LDP. For optical
transitions $|\eta_{j\alpha}^{\rm eff}|/|\eta_{11}^{\rm eff}|\approx
|\eta_{j}^\alpha|/|\eta_{1}^1| = S_j^\alpha \nu_\alpha^{-(1/2)} /
S_1^1 \nu_1^{-(1/2)}$ with $j=\alpha$. The ratio of these parameters
for the highest vibrational mode ($\alpha=10$) is about 3 orders of
magnitude smaller than for mode 1, since ion 10 is only slightly
displaced from its equilibrium position when mode 10 is excited
(compare the discussion in section \ref{Optical}).

We will now investigate how well simultaneous cooling can be done
with the sideband resonances not perfectly superposed.

\subsection{Simultaneous Raman cooling with non-ideal gradient}

\begin{figure}[htbp]
\begin{center}
\includegraphics[width=6cm]{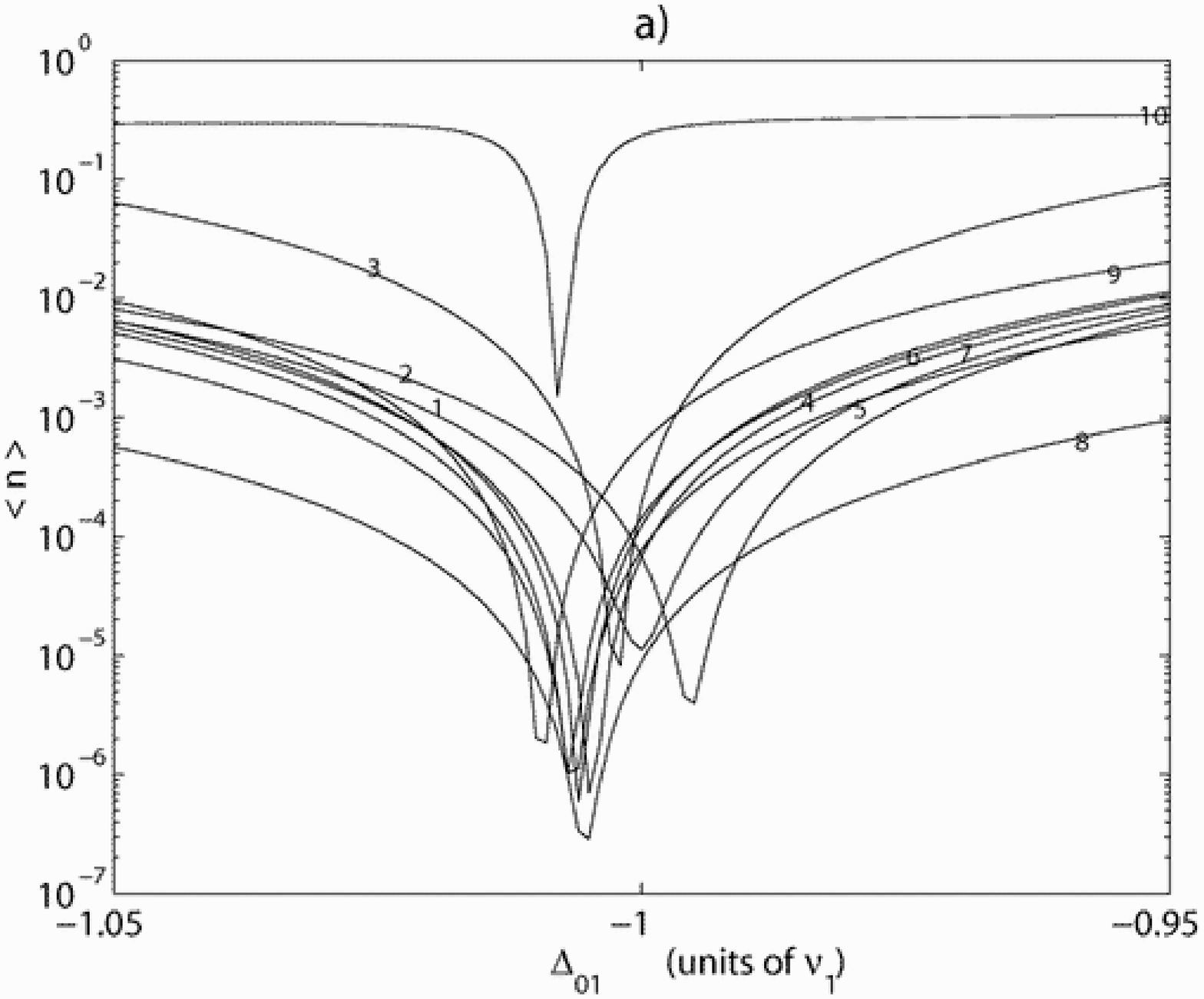}
\includegraphics[width=6cm]{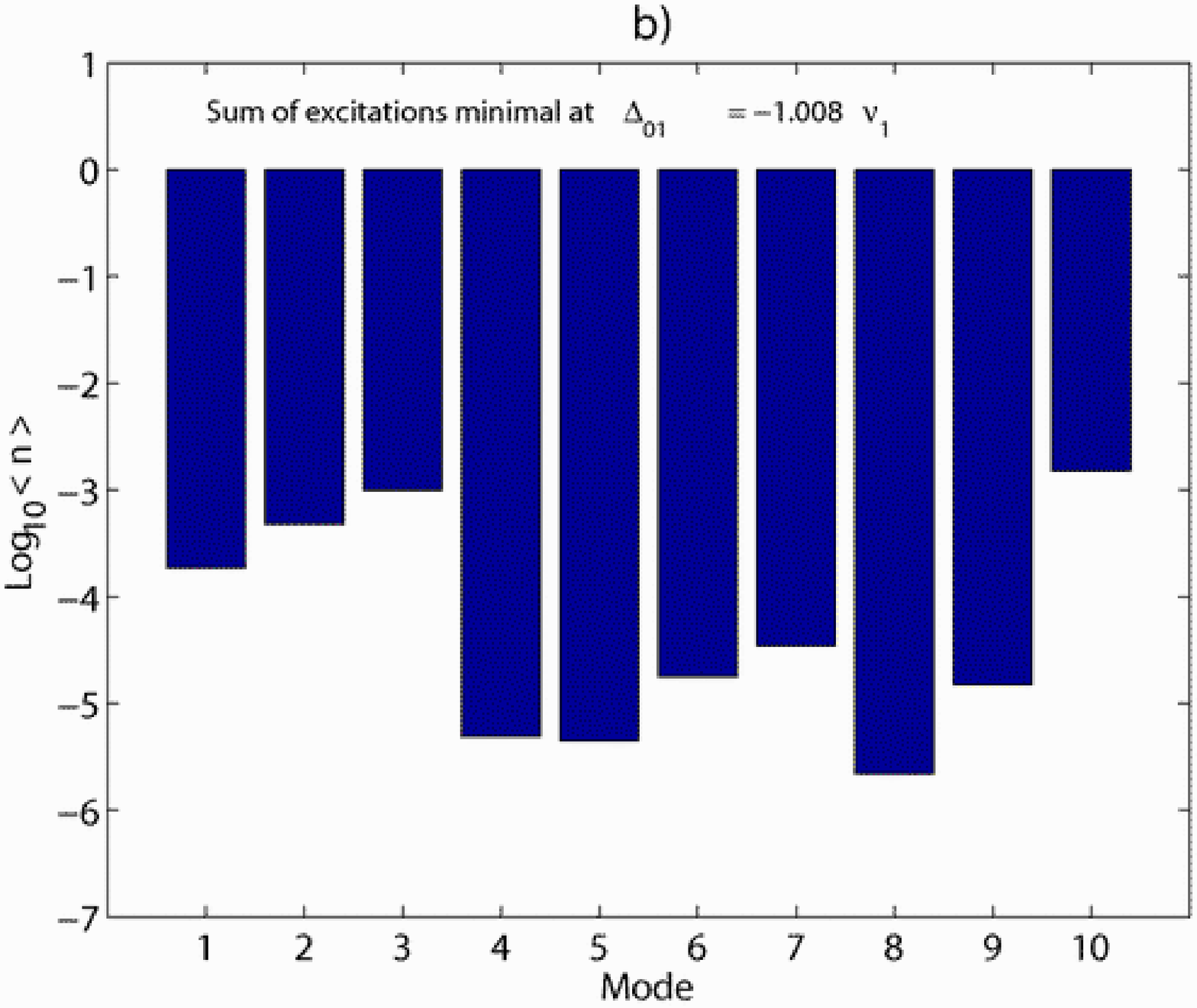}
\caption{Raman sideband cooling a chain of 10 ions in the presence
of a spatially inhomogeneous magnetic field.  The condition
$\omega_1-\nu_1=\ldots =\omega_N-\nu_N$ is {\em approximately}
fulfilled (see Fig.~\ref{FigSideband}). Same parameters as in Fig.
\ref{cool_1} (i.e., $\Omega_{12}=100\times 2\pi$kHz, $\Omega_{01}=5
\times 2\pi$kHz, $\Delta_{12} =-10\times 2\pi$MHz). a) Steady state
vibrational excitation $\langle n^{(\alpha)} \rangle_f$ as a
function of the detuning $\Delta_{01}$ around $\Delta_{01}=-\nu_1$.
b) $\langle n^{(\alpha)} \rangle_f$ for each mode at that detuning
where the sum of the mean vibrational quantum numbers of all ten
modes is minimal.} \label{cool_3}
\end{center}
\end{figure}

\begin{figure}[htbp]
\begin{center}
\includegraphics[width=6cm]{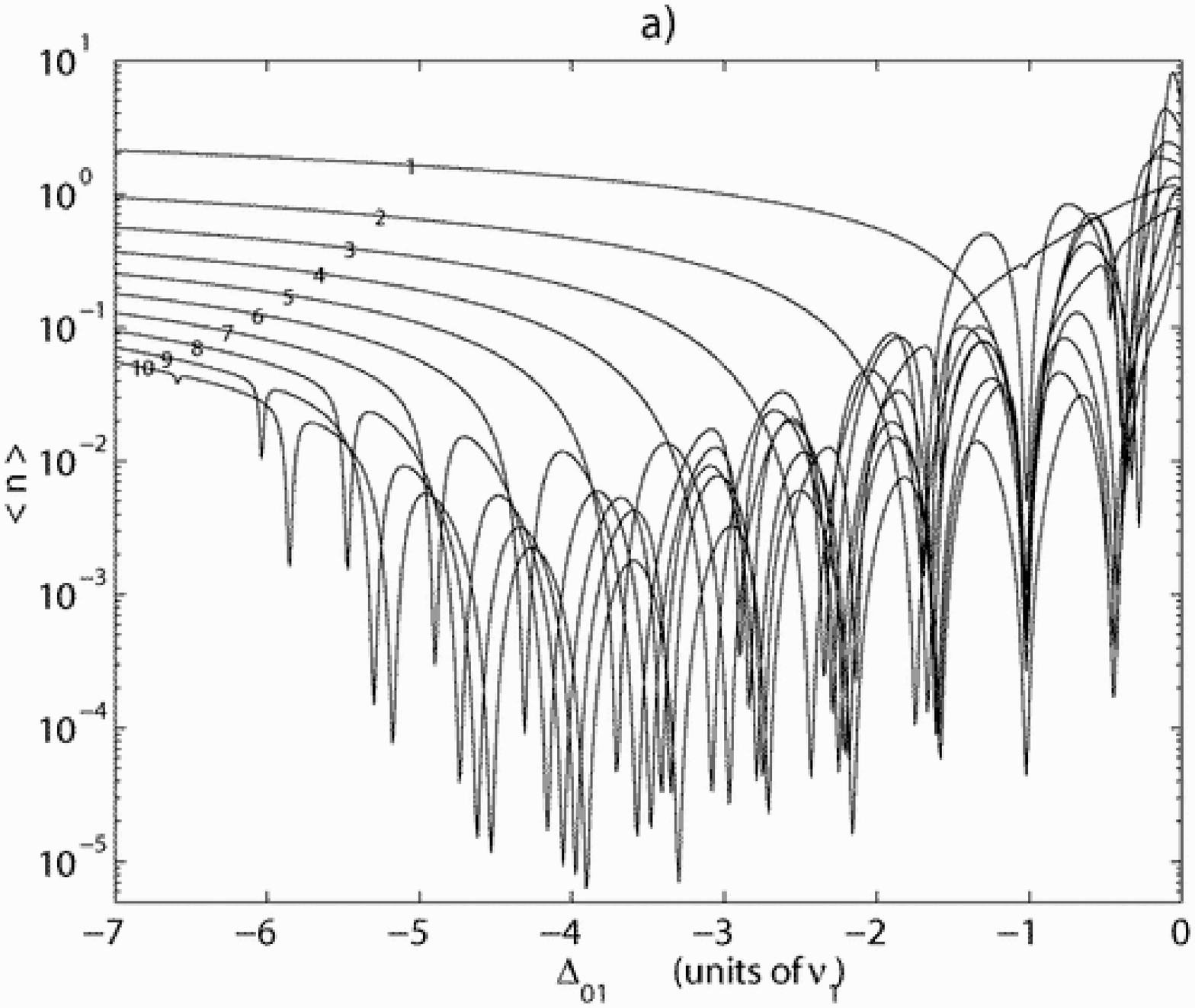}
\includegraphics[width=6cm]{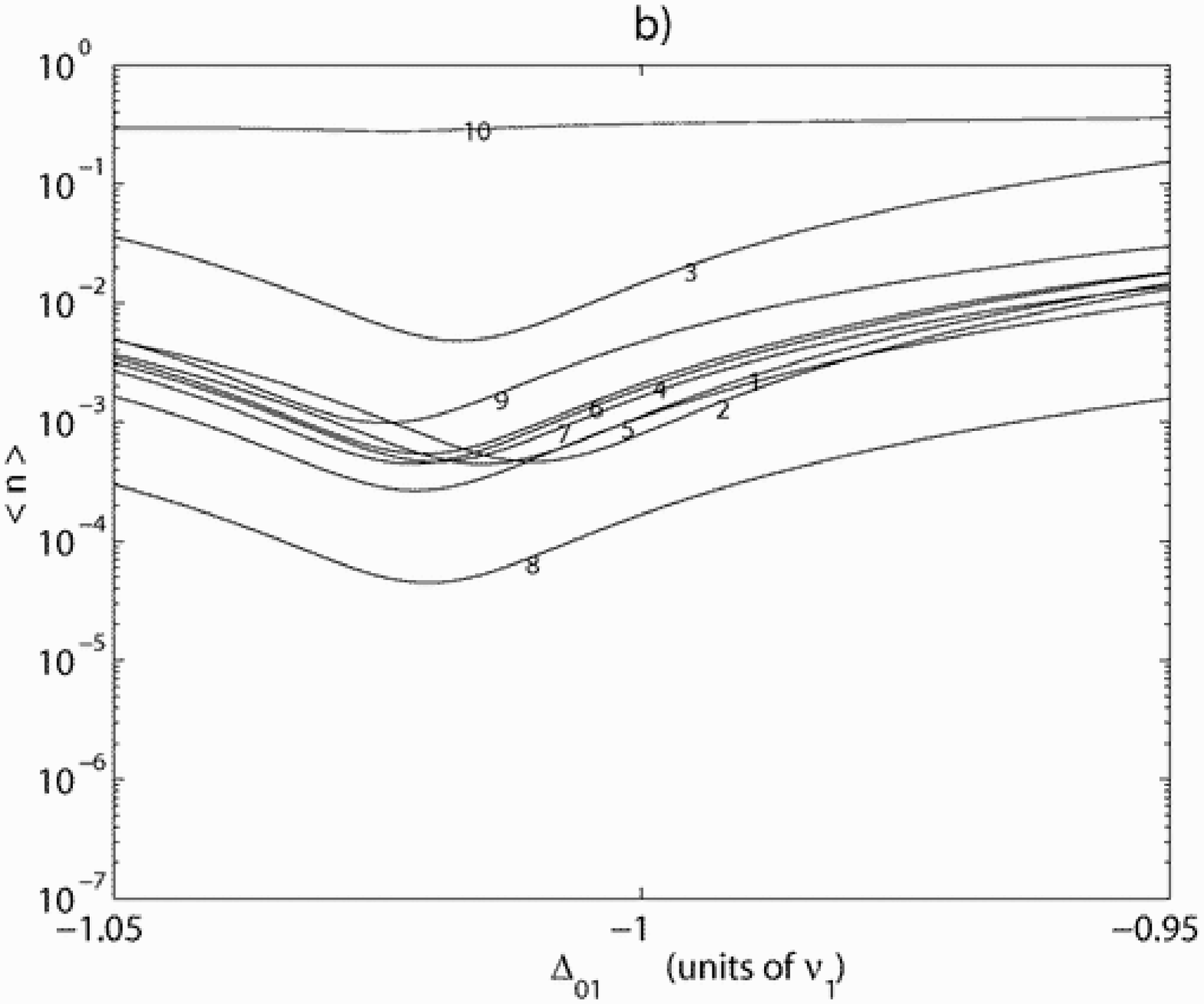}
\includegraphics[width=6cm]{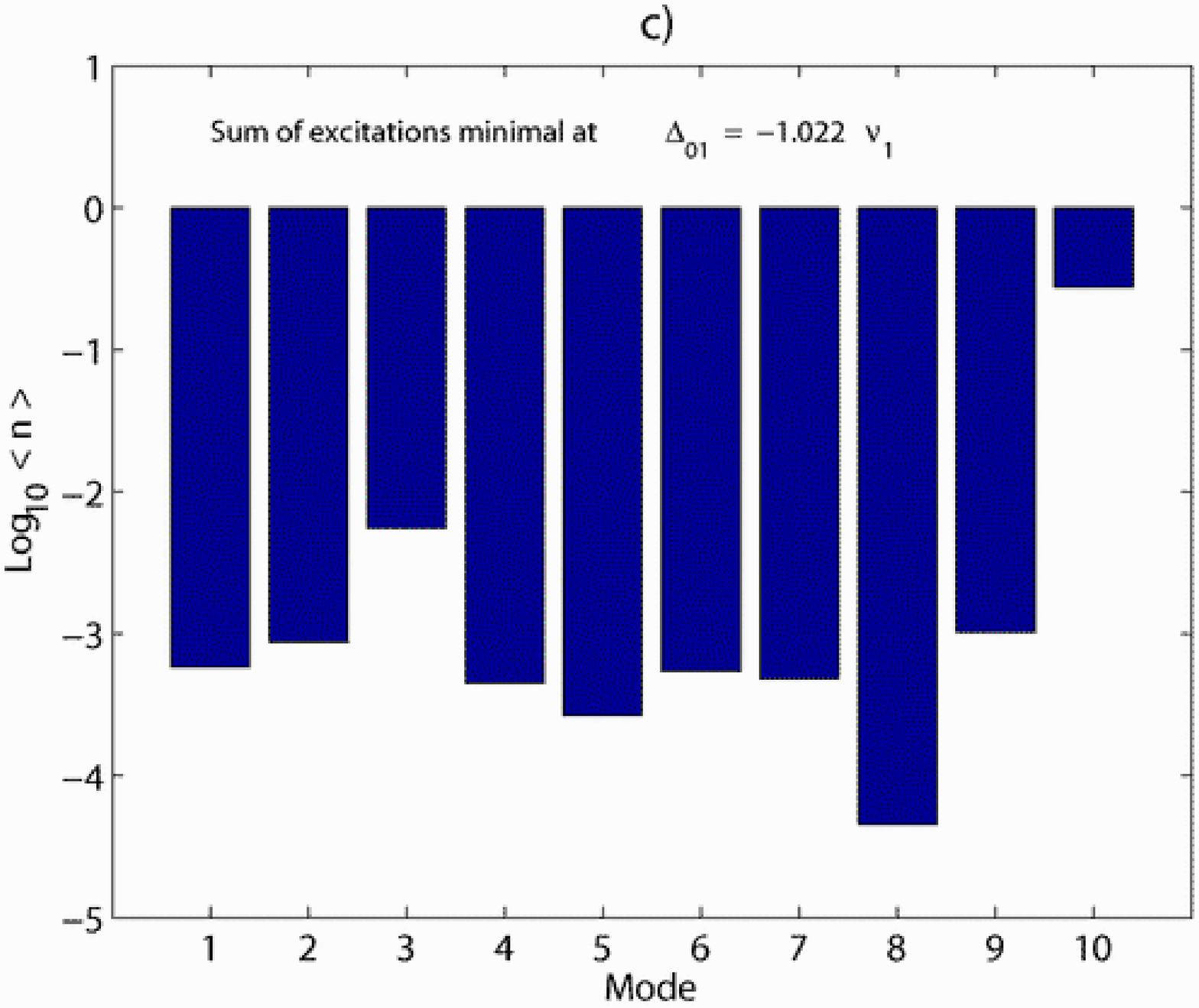}
\includegraphics[width=6cm]{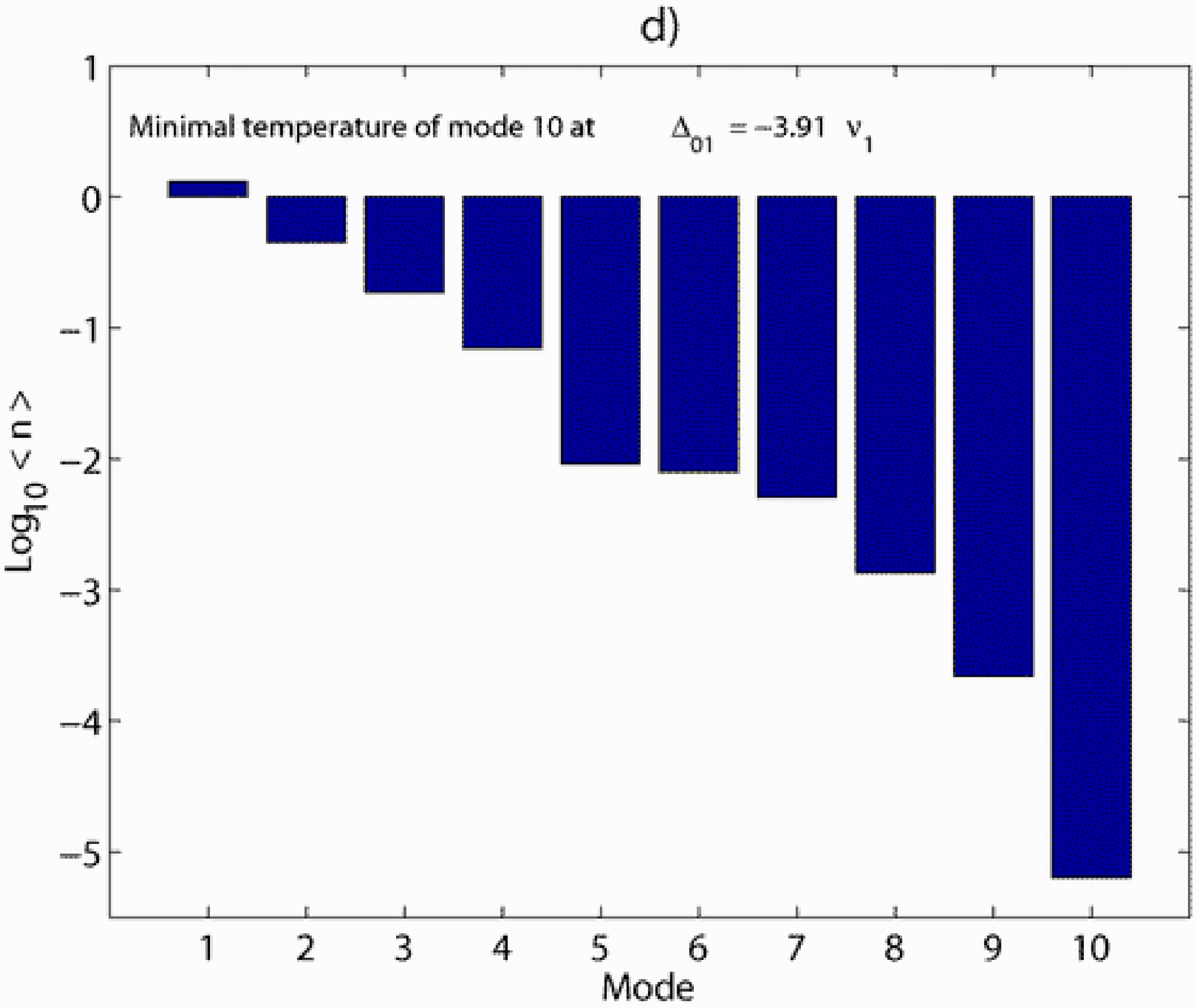}
\caption{Raman sideband cooling a chain of 10 ions in the presence
of a spatially inhomogeneous magnetic field.  The condition
$\omega_1-\nu_1=\ldots =\omega_N-\nu_N$ is {\em approximately}
fulfilled (see Fig.~\ref{FigSideband}). a) $\langle n^{(\alpha)}
\rangle_f$ as a function of $\Delta_{01}$ displaying all first order
red sidebands with all parameters unchanged compared to Fig.
\ref{cool_3} except $\Omega_{12} = 1\times 2\pi$MHz. b) $\langle
n^{(\alpha)} \rangle_f$ as a function of $\Delta_{01}$ in the
vicinity of the common sideband resonance around
$\Delta_{01}=-\nu_1$. The resonances are power broadened and
ac-Stark shifted compared to the ones in Fig. \ref{cool_3}a). c)
Steady state population $\langle n^{(\alpha)} \rangle_f$ for each
mode at that detuning where the sum of the mean vibrational quantum
numbers of all ten modes is minimal (the analogue to Fig.
\ref{cool_3}b except that $\Omega_{12} = 1\times 2\pi$MHz.) d)
$\langle n^{(\alpha)} \rangle_f$ for each mode at that detuning
where one of the modes (here mode 10) reaches the absolute minimum.
Optimal cooling of all vibrational modes is achieved by first
cooling mode 10 at $\Delta_{01}=-3.91\nu_1$ and subsequently modes 1
through 9 at $\Delta_{01}=-1.022\nu_1$.} \label{cool_3_B}
\end{center}
\end{figure}

In Fig. \ref{cool_3}a) the steady state vibrational excitation,
$\langle n^{(\alpha)} \rangle_f $ of a string of 10 ions is
displayed as a function of the detuning of the Raman beams relative
to the resonance frequency $\omega_1$ of ion 1. The Rabi frequencies
and detuning, too, are the same as have been used to generate Fig.
\ref{cool_1}. However, the field gradient that shifts the ions'
resonances is not assumed ideal as in Fig. \ref{cool_1}, instead the
one generated by three single windings as described above (Fig.
\ref{FigGradient}) has been used. Despite the imperfect
superposition of the cooling resonances, low temperatures of all
modes close to their ground state can be achieved as can be seen in
Fig. \ref{cool_3}b). Here, the value of $\langle n^{(\alpha)}
\rangle_f $ for each mode has been plotted at that detuning
$\Delta_{01}=-1.008\nu_1$ where the sum of all excitations is
minimal.

Fig. \ref{cool_3_B}a displays the excitation of each mode over a
wide range of the detuning such that all first order red sideband
resonances are visible. Here, the Rabi frequency $\Omega_{12}$ of
the repump laser has been increased to $1\times 2\pi$MHz as compared
to $100\times 2\pi$kHz in the previous figures. This results i) in
higher final temperatures, and, ii) in broader resonances as is
evident in Fig. \ref{cool_3_B}b and thus makes cooling less
susceptible to errors in the relative detuning between laser light
and ionic resonances.

A higher steady state vibrational excitation due to the larger
intensity of the repump laser is evident in Fig. \ref{cool_3_B}c
where $\langle n^{(\alpha)} \rangle_f$ for each mode is plotted with
the same parameters as in Fig. \ref{cool_3}b, however with
$\Omega_{12}=1\times 2\pi$MHz.

Errors and fluctuations in the relative detuning of the Raman laser
beams driving the sideband transition are expected to be small and
not to affect the efficiency of simultaneous cooling, if the two
light fields inducing the stimulated Raman process are derived from
the same laser source using, for example, acousto-optic or
electro-optic modulators. This is feasible by translating into the
optical domain the microwave or radio frequency that characterizes
the splitting of states $|0\rangle$ and $|1\rangle$. Microwave or rf
signals can be controlled with high precision and display low enough
drift to ensure efficient cooling. If a large enough intensity of
the repump laser is employed, then the steady state vibrational
excitation varies slowly as a function of $\Delta_{01}$ as is
visible in Fig. \ref{cool_3_B}b. Thus, the requirements regarding
both the precision of adjustment and the drift of the source
generating the Raman difference frequency are further relaxed.

It should be noted that efficient cooling does not only occur around
the resonance $\Delta_{01}=-\nu_1$ but also at other values of
$\Delta_{01}$ as can be seen in Fig. \ref{cool_3_B}a. As an example,
Fig. \ref{cool_3_B}d shows $\langle n^{(\alpha)} \rangle_f$ of all
modes at that detuning, $\Delta_{01}=-3.91 \nu_1$ where mode 10
reaches its absolute minimum. At this resonance the red sideband of
the 5th ion corresponding to the 10th mode is driven by the Raman
beams (compare Fig. \ref{scheme}). Note that all other vibrational
modes are also cooled at the same time. Therefore, an efficient
procedure for cooling {\em all} vibrational modes close to their
ground state would be to first tune the Raman beams such that mode
10 is optimally cooled (i.e., $\Delta_{01}=-3.91 \nu_1$, Fig.
\ref{cool_3_B}d), and subsequently set $\Delta_{01}=-1.022 \nu_1$
(Fig. \ref{cool_3_B}c) in order to simultaneously cool modes 1
through 9. This approach is discussed in more detail in section
\ref{SimSeqCooling}.

Initial cooling of vibrational modes often is prerequisite for
subsequent coherent manipulation of internal and motional degrees
of freedom of an ion chain, for example, quantum logic operations.
When implementing quantum logic operations it may not be
advantageous to address all motional sidebands with a single
frequency as is done here for simultaneous sideband cooling. The
magnetic field gradient that superposes the sideband resonances
for cooling may then be turned off adiabatically after initial
Raman cooling of all vibrational modes. This should be done fast
enough not to allow for appreciable heating of the ion string, for
example, by patch fields, and slow enough not to excite
vibrational modes in the process. A lower limit for the time it
takes to ramp up the gradient seems to be $2\pi/\nu_1$. Thus, with
$\nu_1=1\times 2\pi$MHz the additional time needed to change the
field gradient is negligible compared to the time needed to
sideband cool the ion string which for typical parameters takes
between a few hundred $\mu$s and a few ms (compare Fig.
\ref{cool_wB}).

If microwave radiation is used to coherently manipulate internal
and motional degrees of freedom and for quantum logic operations,
it is useful not to turn off the magnetic field gradient, but
instead to ramp up the field gradient to a value where all
coincidences between internal and motional resonances are removed
\cite{Mintert01, Wunderlich03}. Also, a larger field gradient for
quantum logic operations is desirable in this case to have
stronger coupling between internal and external states (i.e., a
larger effective LDP $\eta_{j\alpha}^{\rm eff}$ [eq.
\ref{etaEff}]). The considerations in the previous paragraph
regarding the time scale of change of the field gradient apply
here, too.

For the cooling scheme introduced here to work, the field gradient
has to vary in the axial direction and it remains to be shown in
what follows that this variation is compatible with the neglect of
higher-order terms in the local displacement $q_j$ of ion $j$ and in
$\partial_z B$ in the derivation of the effective Lamb-Dicke
parameter induced by the magnetic field gradient
\cite{Mintert01,Wunderlich02}. Neglecting higher order terms is
justified as long as $|q_j^2\partial_z^2B| \ll |q_j\partial_z B|$.
Using \ref{bz} and $q_j \approx \Delta\!z=\sqrt{\hbar/2m\nu_1}$ this
condition can be written as
\begin{equation}\label{db2}
\frac{2}{\zeta_{j+1}-\zeta_{j-1}} \left|
\frac{(\upsilon_{j+1}-\upsilon_{j})(\zeta_{j}-\zeta_{j-1})}{(\upsilon_{j}-\upsilon_{j-1})(\zeta_{j+1}-\zeta_{j})}
-1 \right|
 \ll \frac{\zeta_0}{\Delta\!z}\; .
 \end{equation}
Considering the region where the second derivative of the magentic
field is maximal ($j=9$) and inserting numbers into relation
\ref{db2} gives for ten ions $0.24\ll 2.9\times 10^{3}
(m/\nu_1)^{\frac{1}{6}}$. The right-hand side of this inequality is
dimensionless if $m$ is inserted in a.m.u., and yields $\approx 500$
for \Yb ions and $\nu_1= 1\times 2\pi$MHz. Hence, relation \ref{db2}
is fulfilled. This can be understood by considering that the typical
distance, $\zeta_0$ over which the gradient has to vary is much
larger than the range of motion, $q_j\approx\Delta\!z$ of an
individual ion, and the approximation of a linear field gradient is
a good one.

\section{Conclusions}
\label{Conclusion}

We have proposed a scheme for cooling the vibrational motion of ions
in a linear trap configuration. Axial vibrational modes are
simultaneously cooled close to their ground state by superimposing
the red motional sidebands in the absorption spectrum of different
ions such that the red sideband of each mode
is excited when driving an internal transition of
the ions with monochromatic radiation. This spectral property is
achieved by applying a magnetic field gradient along the trap axis
shifting individually the internal ionic resonances by a desired
amount.

Exemplary results of numerical simulations for the case of an ion
chain consisting of $N=10$ ions are presented and extensively
discussed. Detailed simulations have also been carried out with
$1<N<10$ and for some values $N>10$. They lead to the same
qualitative conclusion as presented for the case of $N=10$.

Numerical studies show that 
simultaneously Raman cooling all axial
modes is effective for realistic sets of parameters.
These studies
also reveal that using microwave radiation to drive the sideband
transition is not as efficient, due to the relatively small
mechanical effect associated with the excitation of this transition.
The mechanical effect could be enhanced by applying a larger
magnetic field gradients. For simultaneous sideband cooling of all
vibrational modes, however, the gradient is fixed by the requirement
of superimposing the sidebands. Usual sideband cooling on a
hyperfine or Zeeman transition using microwave radiation becomes
possible, if a larger field gradient is used. This may be
particularly useful, if a single ion (or an atom in an optical
dipole trap) is to be sideband cooled using a microwave transition,
or, if the vibrational modes of an ion chain are to be cooled
sequentially using microwave radiation. Moreover, these techniques
could also be implemented for sympathetic cooling of an ion chain,
whereby some modes are simultaneously cooled by addressing ions of
other species embedded in the
chain~\cite{Morigi01,Kielpinski01,Barrett03}.

In conclusion, we have shown that simultaneous sideband cooling with
optical radiation can
be efficiently implemented taking into account experimental
conditions and even for a simple arrangement of magnetic field
generating elements.

\section{Acknowledgements}
We acknowledge financial support by the Deutsche
Forschungsgemeinschaft, Science Foundation Ireland under Grant No.
03/IN3/I397, and the European Union (QGATES,QUIPROCONE).

\section{Appendix}
\label{model}

In this appendix we introduce the hamiltonian and master equation
describing the dynamics discussed in Sec.~\ref{SimCoolingMW}. We
consider a chain consisting of $N$ identical ions aligned along the
$z$-axis, and in presence of a magnetic field $B(z)$. The internal
electronic states of each ion which are relevant for the dynamics
are the stable states $|0\rangle$ and $|1\rangle$ and the excited
state $|2\rangle$. The transitions $|0\rangle\to |1\rangle$,
$|1\rangle\to |2\rangle$ are respectively a magnetic and an optical
dipole transition. We assume that the magnetic moments of
$|0\rangle$ and $|2\rangle$ vanish, while $|1\rangle$ has magnetic
moment $\mu$. Thus its energy with respect to $|0\rangle$ is
shifted proportionally to the field,
$\hbar\omega_0(z)\propto |B(z)|$. The Hamiltonian describing the
internal degrees of freedom has the form:
\begin{equation}
H_{\rm int}=\hbar\sum_j\left(\omega_0(z_j)|1\rangle_j\langle 1|
+\omega_2|2\rangle_j\langle 2|\right)
\end{equation}
where the index $j$ labels the ions along the chain.
The collective excitations of the chain are described by the
eigenmodes at frequency $\nu_1,\ldots,\nu_N$, which are
independent of the internal states. Denoting with
$Q_{\alpha},P_{\alpha}$ the normal coordinates and conjugate
momenta of the oscillator at frequency $\nu_{\alpha}$, the
Hamiltonian for the external degrees of freedom then has  the form
\begin{eqnarray}
H_{\rm mec}
&=& \frac{1}{2m}\sum_{\alpha=1}^N P_{\alpha}^2 \\
   & & + \frac{m}{2}\sum_{\alpha=1}^N \nu_{\alpha}^2
\left[Q_{\alpha} + \frac{\hbar}{2m\nu_{\alpha}^2}\sum_j
       \left.\frac{\partial\omega_j}{\partial z_j}\right|_{z_{0,j}}
|1\rangle\langle 1| S_{j}^{\alpha}\right]^2 \nonumber
\end{eqnarray}
where we have neglected the higher spatial derivatives of the
magnetic field.

Thus, the coupling of the excited state $|1\rangle$ to a spatially
varying magnetic field shifts the center of the oscillators for
the ions in the electronic excited state.\\
The states $|0\rangle$ and $|1\rangle$ are coupled by radiation according to the
Hamiltonian
\begin{equation}
W_{01}=\sum_j\frac{\hbar\Omega_{01}}{2}\left[|1\rangle_j\langle
0| {\rm e}^{-{\rm i}(\omega_{01}t-k_{01}z_j+\psi)}+{\rm
H.c.}\right]
\end{equation}
where the coupling can be generated either by microwave radiation driving the magnetic dipole, or by a pair of Raman lasers. In the first case, $\Omega_{01}$ is the Rabi frequency, $\omega_{01}$ the frequency of radiation and $k_{01}$ the corresponding wave vector. In the case of coupling by Raman lasers, $\Omega_{01}$ is the effective Rabi frequency, $\omega_{01}$ the frequency and $k_{01}$ the resulting wave vector describing the two-photon process.\\
The transition $|1\rangle\to|2\rangle$ is driven below saturation by
a laser at Rabi frequency $\Omega_{12}$, frequency $\omega_{12}$,
and wave vector $k$. The interaction term reads:
\begin{equation}
W_{12}=\sum_j \frac{\hbar\Omega_{12}}{2}\left[|2\rangle_j\langle
1| {\rm e}^{-{\rm i}(\omega_{12}t-kz_j+\psi)} {\rm e}^{{\rm
i}kq_j}+{\rm H.c.}\right]
\end{equation}
where $q_j$ is the displacement of the ion $j$ from the classical
equilibrium position $z_j$.\\
The master equation for the density matrix $\rho$, describing the
internal and external degrees of freedom of the ions, reads:
\begin{equation}
\frac{\partial}{\partial t}\rho=\frac{1}{{\rm i}\hbar}[H,\rho]+
{\cal L}\rho
\end{equation}
where $H=H_0+H_{\rm mec}+W_{01}+W_{12}$, and ${\cal L}\rho$ is the
Liouvillian describing the spontaneous emission processes, i.e.
the decay from the state $|2\rangle$ into the states $|0\rangle$
and $|1\rangle$ at rates $\Gamma_{20}$, $\Gamma_{21}$,
respectively, where $\Gamma=\Gamma_{20}+\Gamma_{21}$ is the total
decay rate. The Liouville operator for the spontaneous decay is
\begin{eqnarray}
& & {\cal L}\rho=-\frac{1}{2}\Gamma\sum_j\Bigl[|2\rangle_j\langle
2|\rho +\rho |2\rangle_j\langle 2|\Bigr]\\
& & +\Gamma_{20}\int_{-1}^1 {\rm d}u{\cal N}(u) {\rm e}^{{\rm
i}ku q_j}|0\rangle_j\langle 2|\rho |2\rangle_j\langle
0|{\rm e}^{-{\rm i}ku q_j} \nonumber\\
& & +\Gamma_{21}\int_{-1}^1 {\rm d}u{\cal N}(u) {\rm e}^{{\rm
i}ku q_j}|1\rangle_j\langle 2|\rho |2\rangle_j\langle
1|{\rm e}^{-{\rm i}ku q_j} \nonumber
\end{eqnarray}
with ${\cal N}(u)$ the dipole pattern for spontaneous emission and $u$ the
projection of the direction of photon emission onto the trap axis.
In order to study the dynamics, it is convenient to move to the
inertial frames rotating at the field frequencies. Moreover, we
apply the unitary transformation \cite{Wunderlich02}
\begin{equation}\label{U}
  U = \exp\left[
            -{\rm i}\sum_{\alpha}\left(
                    \frac{1}{2m\nu_{\alpha}^2}
                    \sum_j\left.\frac{\partial\omega_{0,j}}{\partial z_j}
\right|_{z_{0,n}}
                    |1\rangle_j\langle 1| S_j^{\alpha} \right)P_{\alpha}
           \right]
\end{equation}
We denote with $\tilde{\rho}$ the density matrix in the new
reference frame. The master equation now reads
\begin{equation}
\label{Master:Eq}
\frac{\partial}{\partial t}\tilde{\rho} =\frac{1}{{\rm
i}\hbar}[\tilde{H},\tilde{\rho}]+{\cal L}\tilde{\rho}
\end{equation}
where $\tilde{H}=\tilde{H}_0+\tilde{H}_{\rm mec}+\tilde{W}_{01}+
\tilde{W}_{12}$, and the individual terms have the form:
\begin{equation}
\tilde{H_0}=\hbar\sum_j \left[\delta(z_j)|0\rangle_j\langle
0|+\Delta(z_j)|2\rangle\langle 2|\right]
\end{equation}
with $\delta(z_j)=\omega_{01}-\omega_{0}(z_j)$ and
$\Delta(z_j)=\omega_2-\omega_0(z_j)-\omega_{12}$. The mechanical
energy is
\begin{eqnarray}
\tilde{H}_{\rm mec} &=&\frac{1}{2m}\sum_{\alpha=1}^N P_{\alpha}^2
+ \frac{m}{2}\sum_{\alpha=1}^N \nu_{\alpha}^2 Q_{\alpha}^2 \nonumber\\
&=&\sum_{\alpha}\hbar\nu_{\alpha}\left(a^{\dagger}_{\alpha}a_{\alpha}
+\frac{1}{2}\right)
\end{eqnarray}
where $a^{\dagger}_{\alpha}$, $a_{\alpha}$ are
the creation and annihilation operator, respectively of a quantum
of energy $\hbar\nu_{\alpha}$. The interaction term between the states
$|0\rangle$ and $|1\rangle$ now reads
\begin{equation}
\tilde{W}_{01}=\sum_j\frac{\hbar\Omega_{01}}{2} \left[|1\rangle_j\langle
0|{\rm e}^{{\rm i}(k_{01}z_j-\psi)} \prod_{\alpha}{\rm e}^{-{\rm
i}\tilde{k}_j^{\alpha}P_{\alpha}} +{\rm H.c}\right]
\end{equation}
where
\begin{equation}
\tilde{k}_j^{\alpha}=\frac{\partial\omega_{0,j}}{\partial z_j}
\frac{1}{2m\nu_{\alpha}^2} S_j^{\alpha}
\end{equation}
Thus, the excitation between two states, where the mechanical
potential in one is shifted with respect to the other, corresponds
to an effective recoil, here described by the effective Lamb-Dicke
parameter $\tilde{\eta}_j^{\alpha}=\tilde{k}_j^{\alpha}
\sqrt{\hbar m \nu_{\alpha}/2}$. Finally, the optical pumping between the states
$|1\rangle$ and $|2\rangle$ is given by
\begin{equation}
\tilde{W}_{12}=\sum_j\frac{\hbar\Omega_{12}}{2} \left[|2\rangle_j\langle 1|
{\rm e}^{{\rm i}\chi_j} \prod_{\alpha}{\rm e}^{{\rm i}(\eta^{\rm eff}_{j
\alpha}a^{\dagger}_{\alpha} +\eta^{{\rm eff} *}_{j \alpha}a_{\alpha})} +{\rm
H.c.}\right]
\end{equation}
where $\chi_j$ is a constant phase, that depends on the position
according to
\begin{equation}
\chi_j=kz_j-\psi-\hbar
k\sum_{\alpha}\tilde{k}_j^{\alpha}S_j^{\alpha}/2
\end{equation}
and $\eta^{\rm eff}_{j \alpha}$ is an effective Lamb-Dicke parameter
defined in Eq.~(\ref{etaEff}). The dynamics of the individual modes
are effectively decoupled from the others in the Lamb-Dicke regime,
which holds when the condition $|\eta^{\rm
eff}_{j\alpha}|\sqrt{\langle n^{\alpha}\rangle}\ll 1$ is fulfilled.

The cooling rate and steady state occupation are evaluated for each
mode by using the procedure outlined in~\cite{Marzoli94}. The rates
$A_{\pm}^{\alpha}$, which determine the cooling rate and the steady
state mean occupation  according to Eqs.~(\ref{Gamma})
and~(\ref{nsteady}), are given by
\begin{equation}
A_{\pm}^{\alpha}=2{\rm Re}\{S_{\alpha}(\mp\nu_{\alpha})+D_{\alpha}\}
\end{equation}
where $S_{\alpha}(\mp\nu_{\alpha})$ is the fluctuation spectrum of the dipole force $F_{j,\alpha}$,
\begin{equation}
\label{Snu}
S_{\alpha}(\pm\nu_{\alpha})=
\frac{1}{M\nu_{\alpha}}\sum_j\int_0^{\infty}{\rm d}t{\rm e}^{\pm {\rm i}\nu_{\alpha} t}
{\rm Tr}\{F_{j,\alpha}(t)F_{j,\alpha}(0)\rho_{\rm St}\}
\end{equation}
and $D_{\alpha}$ is the diffusion coefficient due to spontaneous emission,
\begin{equation}
D_{\alpha}=\phi \sum_j\left(|{\eta_{20}}_{\alpha}^j|^2\Gamma_{20}+|{\eta_{12}}_{\alpha}^j|^2\Gamma_{21}\right)\langle 2|\rho_{\rm St}|2\rangle
\end{equation}
where $\phi=\int du{\cal N}(u)u^2$ and here $\phi=2/5$. Here, $\rho_{\rm St}$ is the stationary solution of Eq.~(\ref{Master:Eq}) at {\it zero order} in the Lamb-Dicke parameter, and the dipole force $F_{j,\alpha}$ is defined as
\begin{equation}
F_{j,\alpha}={\rm i}{\eta_{01}}_{\alpha}^j\frac{\Omega_{01}}{2}|1\rangle\langle 0|+
{\rm i}{\eta_{12}}_{\alpha}^j\frac{\Omega_{12}}{2}|2\rangle\langle 1|+{\rm H.c.}
\end{equation}
The fluctuation spectrum and the diffusion are found by evaluating numerically the steady state density matrix. The two-time correlation function in~(\ref{Snu}) is found by applying the quantum regression theorem according to the master equation~(\ref{Master:Eq}) at zero order in the Lamb-Dicke parameter.


\begin{thebibliography}{10}
\bibitem{Footnote:Example}
Strong radial confinement can be achieved in a linear Paul trap
\cite{Ghosh95} by applying
an rf voltage to the quadrupole electrodes while the weaker
trapping potential along the $z-$axis is generated by a dc
voltage. Another possible realization of a trap potential having
the described symmetry can be obtained
with a Penning trap \cite{Ghosh95,Powell02}.
\bibitem{FootnoteB}
The selection rules actually prohibit a direct radiative
transition between $|2\rangle$ and $|0\rangle$, which in this case
correspond to the \Yb states P$_{1/2}, F=0$ and S$_{1/2}, F=0$,
respectively. Here, we assume that a field is present, coupling
P$_{1/2}, F=0$ and P$_{1/2}, F=1$, thus allowing  to have a
transition from P$_{1/2}, F=0$ to S$_{1/2}, F=0$ with adjustable
rate.
\bibitem{footnoteC}
Experimental data on the heating rates of each individual mode for
10 ions seem not to be available. For the purpose of computing the
relevant time scales for simultaneous and sequential cooling
therefore, the heating rates were set to zero in equations
\ref{T_heat_seq} and \ref{T_heat_sim}. For simplicity it was assumed
that $\langle n \rangle_i$ and $\langle n \rangle_f$ take on the
same value for each mode.
\bibitem{Fortagh98} Wires with rather small diameter can be used for these
coils when they are in contact with a suitable heat sink. See, for
instance: J. Fortagh, A. Grossmann, C. Zimmermann, T. W. H\"ansch:
Phys. Rev. Lett. \textbf{81}, 5310 (1998).
\end{thebibliography}
\end{document}